\documentclass[journal]{IEEEtran}
\IEEEoverridecommandlockouts
\usepackage{caption}
\usepackage{cite}
\usepackage{amsmath,amssymb,amsfonts,bm,amsthm}
\usepackage{algorithmic}
\usepackage{graphicx}
\usepackage{textcomp}
\usepackage{xcolor}

\newtheorem{prop}{Proposition}

\usepackage{subfigure}
\usepackage{tcolorbox}
\usepackage{booktabs}
\usepackage{tabularx}
\usepackage{multicol}
\usepackage{multirow}
\usepackage{longtable}
\usepackage{nomencl}
\usepackage{enumitem,kantlipsum}
\usepackage{flushend} 

\makenomenclature

\makeatletter
\newcommand{\pushright}[1]{\ifmeasuring@#1\else\omit\hfill$\displaystyle#1$\fi\ignorespaces}
\newcommand{\pushleft}[1]{\ifmeasuring@#1\else\omit$\displaystyle#1$\hfill\fi\ignorespaces}
\makeatother

\usepackage{etoolbox}
\renewcommand\nomgroup[1]{%
  \item[\bfseries
  \ifstrequal{#1}{P}{Physics constants}{%
  \ifstrequal{#1}{N}{Number sets}{%
  \ifstrequal{#1}{O}{Other symbols}{}}}%
]}

\def\BibTeX{{\rm B\kern-.05em{\sc i\kern-.025em b}\kern-.08em
  T\kern-.1667em\lower.7ex\hbox{E}\kern-.125emX}}
\begin{document}
\begingroup
\allowdisplaybreaks

	\bibliographystyle{IEEEtran}
\title{Uncertainty-Informed Renewable Energy Scheduling: A Scalable Bilevel Framework \\

\author{
\IEEEauthorblockN{Dongwei Zhao, Vladimir Dvorkin, Stefanos Delikaraoglou, Alberto J. Lamadrid L., and Audun Botterud}

\vspace{-5ex}

}

\thanks{The information, data, or work presented herein was funded by the Advanced Research Projects Agency-Energy (ARPA-E), U.S. Department of Energy, under Award Number DE-AR0001277. The views and opinions of authors expressed herein do not necessarily state or reflect those of the United States Government or any agency thereof. Parts of the results have appeared in a conference paper \cite{zhao2022uncertainty}.}

\thanks{Dongwei Zhao and Vladimir Dvorkin are with the Energy Initiative and Laboratory for Information and Decision Systems, Massachusetts Institute of Technology, Cambridge, MA, USA  (e-mail: \{zhaodw, dvorkin\}@mit.edu). 

Stefanos Delikaraoglou and Audun Botterud are with the Laboratory for Information and Decision Systems, Massachusetts Institute of Technology, Cambridge, MA, USA (e-mail: sdelikaraoglou@gmail.com, audunb@mit.edu). 

Alberto J. Lamadrid L. is with the  Department of Economics and Department of Industrial and Systems Engineering, Lehigh University, Bethlehem,
PA, USA  (e-mail: ajlamadrid@lehigh.edu).}

}

\maketitle

\begin{abstract}

This work proposes an uncertainty-informed bid adjustment framework for integrating variable renewable energy sources (VRES) into electricity markets.  This framework adopts a bilevel model to compute the optimal VRES day-ahead bids. It aims to minimize the expected system cost across day-ahead and real-time stages and approximate the cost efficiency of the stochastic market design.  However, solving the bilevel optimization problem is computationally challenging for large-scale systems. To overcome this challenge, we introduce a novel technique based on strong duality and McCormick envelopes, which relaxes the problem to a linear program, enabling large-scale applications. The proposed bilevel framework is applied to the 1576-bus NYISO system and benchmarked against a myopic strategy, where the VRES bid is the mean value of the probabilistic power forecast. Results demonstrate that, under high VRES penetration levels (e.g., 40\%), our framework can significantly reduce system costs and market-price volatility, by optimizing VRES quantities efficiently in the day-ahead market.  Furthermore, we find that when transmission capacity increases, the proposed bilevel model will still reduce the system cost, whereas the myopic strategy may incur a much higher cost due to over-scheduling of VRES in the day-ahead market and the lack of flexible conventional generators in real time.

\end{abstract}




\vspace{-2ex}

\section*{Nomenclature}
\addcontentsline{toc}{section}{Nomenclature}
\subsection{Sets and Indexes}

\begin{IEEEdescription}[\IEEEusemathlabelsep\IEEEsetlabelwidth{$TT$}]

\item[$\Lambda$] Set of transmission lines, indexed by $(n,m)$, where $n/m$ is sending/receiving end of a transmission line
\item[$\Omega$] Set of VRES generation scenarios, indexed by $\omega$
\item[$\mathcal{I}$]  Set of conventional units, indexed by $i$ or $j$
\item[$\mathcal{I}^\text{FS}$] Subset of $\mathcal{I}$ with fast-startup generation units
\item[$\mathcal{I}^\text{SL}$] Subset of $\mathcal{I}$ with slow-startup generation units
\item[$\mathcal{K}$] Set of VRES generation units, indexed by $k$ 
\item[$\mathcal{N}$] Set of grid buses, indexed by $n$
\item[$\mathcal{T}$] Set of market operation time slots, indexed by $t$
\item[$(\cdot)_{n}$] mapping of $(\cdot)$ into the set of buses 
\end{IEEEdescription}

\subsection{Parameters}
\begin{IEEEdescription}[\IEEEusemathlabelsep\IEEEsetlabelwidth{$TTt$}]
\item[$C_{i}$]  Variable generation cost of conventional units 
\item[$C_{i}^0$]  No-load cost of conventional units 
\item[$C_{i}^{\text{SU}}$]  Start-up cost of conventional units 
\item[$C^{\text{sh}}$]  Cost of load shedding (value of lost load) 
\item[$C_{i}^{\text{U/D}}$]  Real-time  up-/downward re-dispatch costs
\item[$\overline{F}_{n,m}$] Transmission line capacities

\item[$L_{n,t}$]  Day-ahead load forecast
\item[$L_{n,t,\omega}$] Real-time demand realization 
\item[$\overline{P}_{i}$] Maximum generation capacity of conventional units
\item[$\underline{P}_{i}$] Minimum generation capacity of conventional units
\item[${R}_{i}^{\text{U/D}}$] Upward/downward ramping capacity of conventional units
\item[$\widetilde{W}_{k,t,\omega}$] Real-time VRES power realization
\item[$\overline{W}_{k}$] VRES power capacity  

\item[$x_{n,m}$] Transmission line reactance 
\end{IEEEdescription}

\subsection{Decision Variables}
\begin{IEEEdescription}[\IEEEusemathlabelsep\IEEEsetlabelwidth{$TTt$}]
\item[$\delta_{n,t}^{\text{DA}}$] Day-ahead voltage angle
\item[$c_{i,t}^\text{DA}$] Day-ahead start-up cost of conventional units
\item[$p_{i,t}^{\text{C}}$] Day-ahead generation of conventional units
\item[$p_{k,t}^{\text{W}}$] Day-ahead generation of VRES units  
\item[$u_{i,t}^\text{DA}$] Day-ahead commitment status of conventional units
\item[$W_{k,t}$] Day-ahead bidding quantity of VRES units

\item[$\delta_{n,t,\omega}^{\text{RT}}$] Real-time voltage angle
\item[$c_{i,t,\omega}^\text{RT}$] Real-time start-up cost of conventional units
\item[$l_{n,t,\omega}^{\text{sh}}$] Real-time  shedding of electrical loads 
\item[$p_{k,t,\omega}^{\text{W,cr}}$] Real-time VRES power curtailment
\item[$r_{i,t,\omega}^{\text{U/D}}$] Real-time up-/downward re-dispatch of conventional units  
\item[$u_{i,t,\omega}^\text{RT}$] Real-time commitment status of conventional units

\end{IEEEdescription}

\section{Introduction}

The appeal for a transition to zero-carbon power systems has motivated the large-scale deployment of variable renewable energy sources (VRES). From 2010 to 2020, the globally installed solar PV capacity increased from about 40 GW to 760 GW, while the wind energy capacity grew from about 200 GW to 740 GW\cite{renreport2021}. In many regions, this transition takes place in restructured electricity markets. For instance, in the USA, New York Independent System Operator (NYISO) is mandated to reach 70\% renewable electricity by 2030 \cite{nyisowinplan} and California Independent System Operator (CAISO) is taking actions to align with the state's policy to supply 100\% zero-emission electricity by 2045  \cite{caisowinplan}.  

The existing electricity market design is tailored to the technical and economic characteristics of dispatchable fossil-fueled generators. However, the increasing levels of VRES pose challenges to achieving market efficiency and reliability due to inherent variability and uncertainty. As a result, there are ongoing discussions around revisiting the electricity market design to address these challenges\cite{zhou2022price}.

Current short-term electricity markets are typically arranged in two stages\cite{fundamentals}: a day-ahead market (DAM) cleared before actual operations to establish initial generation schedules and prices, and a real-time market (RTM)  that runs close to actual delivery to compensate for any imbalances from the day-ahead schedule. However,  in the DAM, the uncertainty information about VRES generations is usually summarized in a single-valued point forecast, typically the conditional expectation of the predicted probability distribution \cite{nyisowindrule}. This myopic approach does not account for the re-dispatch cost under generation uncertainty, thus leading to inefficient coordination between the day-ahead and real-time operations.

{To improve the coordination between day-ahead and real-time stages,  there is extensive literature adopting \textit{stochastic optimization}, \textit{chance-constraint optimization}, and \textit{robust optimization} in the electricity market design to account for uncertainty from VRES sources. The works in \cite{wong2007pricing,zavala2017stochastic, Exizidis2019market, Kazempour2018Market
} cleared the market by means of scenario-based stochastic optimization, which usually guarantees the least system cost in the expected fashion. However, this approach relies on accurate probability distributions of random variables. Also, it is demonstrated that scenario-based stochastic optimization has difficulty in achieving revenue adequacy and cost recovery for each scenario individually \cite{morales2012pricing}.  Some papers \cite{mieth2020risk,dvorkin2019chance} proposed market mechanisms based on the chance-constraint optimization. The chance-constraint optimization can internalize the stochasticity of VRES using statistical moments (e.g., mean and standard deviation). Although the chance constraints can easily control risk tolerance in the market, the underlying assumptions to reformulate chance constraints are restrictive (i.e., they require a linear dependency of optimization variables on random forecast errors) and may lead to sub-optimal, conservative solutions, as shown in \cite{10068745}. The works in \cite{velloso2019two,cobos2018network}  adopted robust optimization to clear the market. Robust optimization seeks a solution that minimizes the worst-case cost,  without explicitly modeling probability distributions of VRES. However, it only optimizes the worst case and may not achieve good performance for other cases. Besides, it is generally challenging to solve robust optimization problems due to its non-convex nature, which makes it difficult to support market operations.}

However, the literature above all cleared the DAM and RTM in a joint manner to determine the DAM schedule, which is not directly compatible with the deterministic two-settlement market structure. To address this issue, Morales \textit{et al.} \cite{morales2014electricity} proposed an adjustment of day-ahead wind-energy quantities using bilevel optimization, which approximates the stochastic market solution within the sequential deterministic energy-only market design. Follow-up work in \cite{viafora2020dynamic,delikaraoglou2019optimal,dvorkin2018setting} extended this bilevel approach to reserve and energy co-optimization.

For practical large-scale power systems, however, solving bilevel problems efficiently is generally difficult due to their non-convex nature. In this work, our bilevel model is bilevel linear programming that has linear objectives and constraints at both lower and upper levels. The works \cite{morales2014electricity, viafora2020dynamic, delikaraoglou2019optimal, dvorkin2018setting} solved bilevel linear programming for market clearing by formulating mixed-integer linear programming (MILP) problems based on Karush--Kuhn--Tucker (KKT) conditions. However, these methods were only tested on small-scale systems, such as a 24-bus system \cite{morales2014electricity}. For large-scale systems, MILP problems can still be challenging to solve, making it difficult to support practical market operations. For instance, on the NYISO system with over one thousand buses, the reformulated MILP problem cannot compute results within two hours for economic dispatch problems \cite{zhao2022uncertainty}. Therefore, alternative methods are necessary to address the computational complexity of bilevel programming for larger-scale systems.

While bilevel linear programming has been extensively studied, there is no guarantee of efficient approaches for large-scale problems \cite{dempe2020bilevel}. Among proposed techniques, enumerative methods search for extreme points with iterative linear programming (LP) problems \cite{bialas1982two, dempe1987simple, zhao2019virtual}. Other heuristic algorithms lack theoretical guarantees and require solving many intermediate optimization problems \cite{stojiljkovic2017bi, kuo2009application}. In the domain of energy and power systems, the most common approach involves replacing the lower-level problem with KKT conditions \cite{dempe2020bilevel,salmeron2009,viafora2020dynamic, delikaraoglou2019optimal, dvorkin2018setting}. This reformulated problem can be solved using commercial MILP solvers \cite{ kleinert2021}, branch-and-bound algorithms \cite{bard1990branch}, Benders decomposition \cite{salmeron2009}, or penalty function methods \cite{white1993penalty}. However, at scale, these methods either introduce a large number of binary variables or require many iterations of large-scale optimization problems, making them computationally expensive. 

One alternative method for bilevel problems is to replace complementarity conditions in KKT conditions with equivalent strong-duality conditions \cite{boyd2004convex}. However, the strong-duality constraint will introduce non-convex bilinear items. In power and energy research, Zhou \textit{et al.}\cite{zhou2022storage} and Yang \textit{et al.}\cite{yang2022optimal} discretized variables to transform bilinear items of strong-duality conditions in bilevel pricing problems, which creates a large number of binary variables.  Cerisola \textit{et al.} utilized McCormick envelopes \cite{mccormick1976computability} to relax bilinear functions in hydrothermal models \cite{cerisola2012stochastic}.

In this work, we combine the strong-duality condition and McCormick envelope together on the bilevel LP problem, which relaxes it into an LP problem. We demonstrate that this approach is scalable for large-scale systems, while it can achieve significant economic benefits. We present the technical contributions of this work as follows:

    \textit{1)~Bilevel optimization of VRES bids for unit-commitment problems:} {We present a bilevel-dispatch framework (\textit{BiD})  to adjust VRES bidding quantities in the day-ahead unit commitment (UC) problem, which can guarantee the minimum expected system cost in the
context of a deterministic two-settlement market structure.} Relative to \cite{morales2014electricity,viafora2020dynamic,delikaraoglou2019optimal,dvorkin2018setting}, which primarily focuses on the European market settlements,  we include the UC constraints and fit the framework into the scope of US electricity markets. We compare \textit{BiD} with two benchmarks: (i) a myopic-dispatch strategy (\textit{MyD}) where VRES producers bid the forecast (expected) value in the DAM. (ii) a stochastic-dispatch framework (\textit{StD})  that jointly optimizes DAM and RTM, and attains the least expected cost.  
     
     \textit{2)~Efficient algorithm for practical large-scale systems:} Solving large-scale bilevel optimization problems is computationally expensive. The conventional method of reformulating a MILP problem based on  KKT conditions cannot solve the large-scale NYISO system within two hours for economic dispatch problems \cite{zhao2022uncertainty}. To address this scalability issue,  we propose a new approach based on the strong-duality condition and McCormick-envelope relaxation. We design proper relaxation bounds for McCormick envelopes. This method is computationally efficient for the NYISO system as it only requires solving a linear program. Meanwhile, it can achieve significant economic benefits compared with \textit{MyD}.

    \textit{3)~Large-scale application:} We conduct numerical studies on a 1576-bus NYISO system. Under a high VRES penetration level (e.g., 40\%) and current generator settings (73\% fast-startup generation capacity) in NYISO, the results demonstrate significant economic benefits of \textit{BiD} compared to \textit{MyD}. By efficiently optimizing  VRES bidding quantities in DAM, \textit{BiD} can (i) substantially reduce the expected system cost (by over 20\%), (ii) decrease the volatility of market prices and VRES producers'  revenues, and (iii) effectively prevent negative market revenues for VRES producers. Furthermore,  increasing transmission capacities may incur a much higher system cost under \textit{MyD}  than \textit{BiD}. The reason is that excessive renewable energy is scheduled in DAM without adequate flexible generators in RTM to address uncertainty. Notably, these benefits depend upon the penetration levels of VRES  and generator flexibility. For instance, if all generators can start up fast in RTM, \textit{MyD} can achieve similar economic benefits to \textit{BiD} and \textit{StD}. 

{Notably,  this work focuses on the system perspective, which establishes benchmark bidding quantities in a centralized way by minimizing the expected system cost. We expect that the benchmark bidding quantities can also serve as a tool to guide, monitor, or regulate the bidding strategies of private suppliers. For example, by comparing the benchmark results with the forecast values, a risk score can be developed to adjust and guide how much suppliers should bid in the market. We next discuss two important future extensions for this work.}

 {First, one future direction is to compare this centralized manner with the case that suppliers make bidding strategies by themselves.
We briefly discussed two possible cases: price-taking and price-making bidding strategies. If the private suppliers just make decisions as price-takers based on the  DAM and RTM price estimation, one of our previous papers \cite{Marija2023dynamic} shows that under certain assumptions, the profit-seeking suppliers should either bid zero or maximum wind quantity in the DAM. This result may not be efficient from the system perspective.
To reach the system optimum in a decentralized way,  the impact of bids on prices needs to be considered in the suppliers' decisions. It is necessary to adopt a dynamic and smart framework to coordinate and guide private suppliers to make decisions. One possible direction is to design a decentralized algorithm or incentive for suppliers based on the centralized bilevel model.}

 {Second, although we use \textit{stochastic optimization} to formulate the bilevel framework in a risk-neutral way, we can also extend this work by adopting different risk-aversion approaches to deal with uncertainty. For example, we can modify the objective in the upper level based on the \textit{conditional value of risk (CVaR)}. In addition, we can impose  chance constraints into upper and lower levels to formulate a \textit{chance-constraint optimization} problem. We can set up an uncertainty set and optimize the worst scenario in the upper-level objective using \textit{robust optimization}. In these cases, the risk-aversion factor can be reflected while the deterministic two-settlement market structure can still be maintained. We leave these risk-aversion extensions as future work.}

The rest of the paper is organized as follows. We first introduce the standard deterministic market-clearing model in Section \ref{sec:market}. Then, we formulate a bilevel framework for VRES quantity adjustments in Section \ref{section:bilevel} and present the solution method in Section \ref{section:solution}. Finally, we  present the results of different case studies and conclude in Sections \ref{section:simulation} and \ref{section:conclusion}.

\section{Two-settlement market clearing}\label{sec:market}
This section provides optimization models for the  day-ahead and real-time unit commitment market dispatch.  In terms of modeling assumptions, the network topology is included considering linear DC power flows.\footnote{ {The Independent System Operators (ISOs) in the US leverage the DC Optimal Power Flow (DCOPF)  formulation to settle and clear the DAM and RTM \cite{cali2021optimization}.  More detailed feasibility checks (e.g. with AC models) can still be performed outside the market clearing formulation, similar to the practice in current electricity markets.}} We assume a linear operation cost for conventional generation units and zero operation cost for VRES.\footnote{ {We use the synthetic data of NYISO systems developed by \cite{greene2022}, which model linear operation costs of thermal units. Our model can easily be generalized to step-wise cost curves of thermal units since the DAM and RTM clearing models still exhibit linear structures. Such a format of step-wise cost curves is adopted in the NYISO \cite{nyisobidrule}. }}
The system demand is inelastic with a large value of lost load (VoLL).\footnote{ { We leave demand response modeling and market participation
to future work. In fact, we can easily generalize our model to incorporate step-wise demand curves to provide
a more detailed representation of load reductions, where the DAM and RTM clearing models still exhibit linear structures. Demand curves can indicate the elasticity of demand,
e.g., a low cost of load reductions shows high elasticity while a high cost shows low elasticity.}} To maintain convexity in the market-clearing models, we relax binary commitment decisions and build a relaxed version of the UC problem \cite{kazempour2017value}\cite{kasina2014comparison}.\footnote{In Appendix \ref{app:uc}, we show that this relaxation well approximates the binary setting, where the non-discrete UC solutions in the range $(0,1)$ make up less than 1\% of the total number of UC solutions.}  We differentiate the fast and slow start-up generators, and omit minimum on- and
off-times for simplicity. For system uncertainty,  we use the scenario set $\Omega$ to model the discrete probability distribution of real-time VRES  generations and system demands.

\subsection{Day-Ahead Market} 
The day-ahead market-clearing problem takes the form:
\begin{subequations}\label{prob:DA}
\allowdisplaybreaks
\begin{align}
   \underset{\Phi^{\text{DA}}}{\min} ~&f^{\text{DA}}(\Phi^{\text{DA}}):= 
        \sum_{i \in \mathcal{I}}\sum_{t\in\mathcal{T}} C_{i}  \cdot p_{i,t}^{\text{C}} +u_{i,t}^\text{DA}\cdot C_i^0+c^{\text{DA}}_{i,t} \\
         \text{s.t.} 
	~&\sum_{i \in \mathcal{I}_{n}}p_{i,t}^{\text{C}} + \sum_{k \in \mathcal{K}_{n}}p_{k,t}^{\text{W}} - L_{n,t} \notag\\ 
        &\hspace{1ex} -\sum_{m:(n,m)\in\Lambda}\frac{\delta_{n,t}^{\text{DA}}-\delta_{m,t}^{\text{DA}}}{x_{n,m}} = 0 
   :\lambda_{n,t}^b,\label{eq:dabalance}\\
	&-\overline{F}_{n,m}\leqslant \frac{\delta_{n,t}^{\text{DA}}-\delta_{m,t}^{\text{DA}}}{x_{n,m}} \leqslant \overline{F}_{n,m}:\underline{\lambda}_{n,m,t},\overline{\lambda}_{n,m,t},\label{eq:daline}\\
    &\pushright{\forall n\in\mathcal{N}, \forall (n,m) \in \Lambda,\forall t \in \mathcal{T},}\nonumber\\
    &0 \leqslant p_{k,t}^{\text{W}} \leqslant {{W}_{k,t}}:\underline{\lambda}_{k,t}^\text{W}, \overline{\lambda}_{k,t}^\text{W}, \label{eq:daw}\\
    &\pushright{\forall k\in \mathcal{K},\forall t \in \mathcal{T}},\nonumber\\
    &u_{i,t}^{\text{DA}}\cdot \underline{P}_{i}\leqslant p_{i,t}^{\text{C}} \leqslant u_{i,t}^{\text{DA}}\cdot \overline{P}_{i}:\underline{\lambda}_{i,t}^\text{C}, \overline{\lambda}_{i,t}^\text{C},\label{eq:dac}\\
    & 0\leqslant u_{i,t}^{\text{DA}} \leqslant 1:\underline{\lambda}_{i,t}^\text{U}, \overline{\lambda}_{i,t}^\text{U},\label{eq:dauc}\\
    & C_i^{\text{SU}}\cdot (u_{i,t}^{\text{DA}}-u_{i,t-1}^{\text{DA}})\leqslant c^{\text{DA}}_{i,t}:{\lambda}_{i,t}^{\text{S}_1},\label{eq:dasuc1}\\
    & 0\leqslant c^{\text{DA}}_{i,t}:{\lambda}_{i,t}^{\text{S}_2},\label{eq:dasuc2}\\
    & p_{i,t}^\text{C}-p_{i,t-1}^\text{C} \geqslant - u_{i,t-1}^{\text{DA}}\cdot R_i^\text{D} : \underline{\lambda}_{i,t}^\text{R},   \label{eq:darp_a}\\
    &p_{i,t}^\text{C}-p_{i,t-1}^\text{C} \leqslant u_{i,t}^{\text{DA}}\cdot R_i^\text{U}: \overline{\lambda}_{i,t}^\text{R},\label{eq:darp_b}\\
    &\pushright{\forall i \in  \mathcal{I}, \forall t \in \mathcal{T},} \nonumber
	\end{align}
\end{subequations}
where the total dispatch cost $f^{\text{DA}}$ includes the variable, no-load, and startup costs of conventional generation units. 
The decision variable set $\Phi^{\text{DA}}$ comprises the day-ahead decisions for each conventional unit, i.e., $p_{i,t}^{\text{C}},u_{i,t}^{\text{DA}}, c_{i,t}^\text{DA}$, and each VRES unit, i.e., $p_{k,t}^{\text{W}}$, as well as the
voltage angles $\delta_{n,t}^{\text{DA}}$ for each bus. Constraint \eqref{eq:dabalance} guarantees the day-ahead power balance at each bus. Power flow constraints \eqref{eq:daline} are enforced to satisfy transmission limits. The generation limits of renewable and conventional power producers are enforced using constraints \eqref{eq:daw} and \eqref{eq:dac}, respectively, where the latter accounts for the UC decisions. A final set of constraints \eqref{eq:dauc}--\eqref{eq:darp_b} model the relaxed UC decisions, UC costs, and ramping limits.  Note that the subscript $t=0$ denotes the initial states of variables before the operation horizon $\mathcal{T}$.


For VRES in constraint \eqref{eq:daw}, we define the VRES quantity offer $\bm{W}=(W_{k,t},\forall k\in \mathcal{K}, \forall t\in \mathcal{T})$. We denote by $\mathcal{X}^{\text{DA}}(\bm{W})$ the constraint set constructed by  \eqref{eq:dabalance}-\eqref{eq:darp_b} parameterized by VRES offer $\bm{W}$.  We list the dual variables after the column signs in \eqref{eq:dabalance}-\eqref{eq:darp_b}, and gather them in vector $\bm{\lambda}$. Finally, we will denote the optimal day-ahead solution by $\Phi^{\text{DA}\star}$.

In the baseline dispatch, $\bm{W}$ is set to the mean value of the VRES forecast. Later in Section \ref{section:bilevel}, we will introduce the bilevel model which optimizes $\bm{W}$ to achieve better coordination between the day-ahead and real-time markets.

\subsection{Real-Time Market}

Closer to real-time operations, any deviation from the  day-ahead schedule $\Phi^{\text{DA}\star}$ is compensated for by balancing actions. For a specific realization $\omega\in \Omega$ of renewable generation {$\widetilde{W}_{k,\omega,t}$} and demand {${L}_{k,\omega,t}$}, the system operator decides the optimal re-dispatch by minimizing the re-dispatch cost $f_\omega^{\text{RT}}$, which includes the upward/downward adjustment  cost of conventional generators, startup cost and fixed operational cost  of new online generators, and the cost of lost load: 
\begin{subequations}\label{prob:RT}
\allowdisplaybreaks
	\begin{align}
&\underset{\Phi_\omega^{\text{RT}}}{\min}~  f_\omega^{\text{RT}}(\Phi_\omega^{\text{RT}},\Phi_\omega^{\text{DA}\star}):=\sum_{i \in \mathcal{I}}\sum_{t\in \mathcal{T}} \Big( C_{i}^\text{U} r_{i,t,\omega}^{\text{U}}- C_{i}^\text{D} r_{i,t,\omega}^{\text{D}} \notag \\
&+C_i^0\cdot (u_{i,t,\omega}^{\text{RT}}\!-u_{i,t}^{\text{DA}\star})\!+\!c_{i,t,\omega}^{\text{RT}}\Big) +\sum_{n \in \mathcal{N}}\sum_{t\in \mathcal{T}} C^{\text{sh}} l_{n,t,\omega}^{\text{sh}} \\
&	\text{s.t.}~  \sum_{i \in \mathcal{I}_{n}} \left( p_{i,t}^{\text{C}\star}+r_{i,t,\omega}^{\text{U}}-r_{i,t,\omega}^{\text{D}} \right)\!+\! \sum_{k \in \mathcal{K}_{n}}\!\left(\widetilde{W}_{k,t,\omega} -  p_{k,t,\omega}^{\text{W,cr}} \right) \notag  \\
&\hspace{6ex}- \hspace{-1ex}
\sum_{m:(n,m)\in\Lambda} \hspace{-2ex}\frac{\delta_{n,t,\omega}^{\text{RT}}-\delta_{m,t,\omega}^{\text{RT}}}{x_{n,m}} = L_{n,t,\omega}-l_{n,t,\omega}^{\text{sh}}, \label{eq:rtbalance}\\
&0 \leqslant l_{n,t,\omega}^{\text{sh}} \leqslant L_{n,t,\omega}, \label{eq:rtlost}\\
&\pushright{\forall n \in \mathcal{N},\forall t \in \mathcal{T},}\nonumber \\
& -\overline{F}_{n,m}\leqslant \frac{\delta_{n,t,\omega}^{\text{RT}}-\delta_{m,t,\omega}^{\text{RT}}}{x_{n,m}} \leqslant \overline{F}_{n,m},\label{eq:rtline} \\
&\pushright{\forall (n,m) \in \Lambda, \forall t \in \mathcal{T}},\nonumber\\
& u_{i,t}^{\text{DA}\star} \leqslant u_{i,t,\omega}^{\text{RT}} \leqslant 1,\quad u_{j,t}^{\text{DA}\star} = u_{j,t,\omega}^{\text{RT}}, \label{eq:rtfs}\\
&\pushright{\forall i \in  \mathcal{I}^{\text{FS}}, \forall j \in  \mathcal{I}^{\text{SL}}, \forall t \in \mathcal{T},}\nonumber\\
& u_{i,t,\omega}^{\text{RT}}\cdot \underline{P}_{i}\leqslant p_{i,t}^{\text{C}\star}+ r_{i,t,\omega}^{\text{U}}-r_{i,t,\omega}^{\text{D}} \leqslant  u_{i,t,\omega}^{\text{RT}}\cdot \overline{P}_{i}, \label{eq:rtup}\\
&c_{i,t,\omega}^{\text{RT}}+c_{i,t}^{\text{DA}\star}\geqslant C_i^{\text{SU}}\cdot (u_{i,t,\omega}^{\text{RT}}-u_{i,t-1,\omega}^{\text{RT}}),\label{eq:rtsuc1}\\
&-u_{i,t-1,\omega}^{\text{RT}}\cdot R_i^\text{D} \leqslant p_{i,t}^{\text{C}\star}+r_{i,t,\omega}^{\text{U}}-r_{i,t,\omega}^{\text{D}}\notag \\
&\hspace{6ex}-(p_{i,t-1}^{\text{C}\star}+r_{i,t-1,\omega}^{\text{U}}\!-\!r_{i,t-1,\omega}^{\text{D}}) \leqslant u_{i,t,\omega}^{\text{RT}}\!\cdot \!R_i^\text{U},\label{eq:rtrp}\\
&c_{i,t,\omega}^{\text{RT}}\geqslant 0,~r_{i,t,\omega}^{\text{U}}\geqslant0, ~r_{i,t,\omega}^{\text{D}}\geqslant0,\label{eq:rtsuc2}\\
&\pushright{\forall i \in \mathcal{I},\forall t \in \mathcal{T}},\nonumber\\
& 0 \leqslant p_{k,t,\omega}^{\text{W,cr}} \leqslant \widetilde{W}_{k,t,\omega},\label{eq:rtw}\\
&\pushright{\forall k \in \mathcal{K},\forall t \in \mathcal{T}.}\nonumber
\end{align}
\end{subequations}
Here the decision variable set $\Phi_\omega ^{\text{RT}}$ includes the re-dispatch variables of each conventional unit, i.e., ($r_{i,t,\omega}^{\text{U}}$, $r_{i,t,\omega}^{\text{D}}$,  $u_{i,t,\omega}^{\text{RT}}$, $c_{i,t,\omega}^{\text{RT}}$), curtailment of each VRES producer $p_{k,t,\omega}^{\text{W,cr}}$, and lost load $l_{n,t,\omega}^{\text{sh}}$,  as well as the  voltage angles $\delta_{n,t,\omega}^{\text{RT}}$ at each bus. Constraint \eqref{eq:rtbalance} ensures the power balance.  Constraints  \eqref{eq:rtlost} and \eqref{eq:rtline} give shedding load feasibility and transmission limits, respectively. The first entry in constraint \eqref{eq:rtfs} allows the fast startup generators that are not committed in the DAM to be dispatched in the RTM. On the other hand, the commitment status of slow-startup generators in the RTM remains the same as in the DAM as per the second entry in \eqref{eq:rtfs}. Constraint \eqref{eq:rtup} is the capacity limit on the re-dispatch of conventional generators. Constraint \eqref{eq:rtsuc1} yields the startup cost. Constraints \eqref{eq:rtrp}, \eqref{eq:rtsuc2}  and \eqref{eq:rtw} ensure the ramping-rate limits, the feasibility of upward/downward adjustment of conventional generators, and real-time VRES curtailment, respectively. 

We denote by $\mathcal{X}_{\omega}^{\text{RT}}(\Phi ^{\text{DA}})$ the constraint set constructed by  \eqref{eq:rtbalance}-\eqref{eq:rtw}, which is coupled with day-ahead schedule $\Phi ^{\text{DA}}$. Note that both optimization problems for the day-ahead and real-time dispatch are LP problems.

\section{Uncertainty-informed quantity adjustments}\label{section:bilevel}
\vspace{-0.5ex}
Building on the sequential DAM and RTM problems \eqref{prob:DA}--\eqref{prob:RT},  we develop a bilevel-dispatch framework (\textit{BiD}) to factor in the uncertainty of renewable generation and reveal the optimal quantity offers of VRES producers, which improves the coordination between the two markets in terms of the dispatch cost. We also introduce two benchmark  dispatch frameworks to assess the performance of the proposed framework.

\subsection{Bilevel Optimization framework}

The framework \textit{BiD} is based on bilevel optimization problems. In the upper level, the system operator decides the day-ahead VRES bidding quantities $\bm{W}$. In the lower level, the day-ahead schedule $\Phi^{\text{DA}}$ is optimized for a fixed value $\bm{W}$ provided from the upper problem. Sourcing the optimal day-ahead dispatch $\Phi^{\text{DA}}$ from the lower-level problem, the system operator  in the upper level also optimizes the real-time re-dispatch $\Phi_{\omega}^{\text{RT}}$ for different uncertainty realizations. The overall goal is to minimize the total system cost, i.e. the sum of DAM and expected RTM costs. The bilevel problem takes the following form:

\begin{tcolorbox}[standard jigsaw,opacityback=0]
\vspace{-1ex}
\textbf{Problem \textit{BiD}: Bilevel optimization problem for day-ahead VRES quantity offer}
\begin{subequations} \label{prob:bilevel_clearing}	
\allowdisplaybreaks
	\begin{align*}
\hspace{-2ex}S^{\text{BiD}}:=\underset{\Phi^{\text{RT}}\bigcup \bm{W}}{\text{min}} ~&
		f^{\text{DA}}(\Phi^{\text{DA}\star}) + \mathbb{E}_{\omega\in \Omega}\left[f_\omega^{\text{RT}}(\Phi_\omega^{\text{RT}},\Phi_\omega^{\text{DA}\star})\right] \\	
	\text{s.t.} ~& \Phi_\omega^{\text{RT}}\in \mathcal{X}_\omega^{\text{RT}}(\Phi^{\text{DA}\star}),~\forall \omega \in \Omega,\\
		&\Phi^{\text{DA}\star}  \in \text{arg}
		\left\{\!\begin{aligned}
			\underset{\Phi^{\text{DA}}}{\text{min}} ~&f^{\text{DA}}(\Phi^{\text{DA}})\\
			\text{s.t.} ~& \Phi^{\text{DA}}\in \mathcal{X}^{\text{DA}}(\bm{W})\\
		\end{aligned}\right\}. \label{eq:LLD}
	\end{align*}
\end{subequations}
\vspace{-2ex}
\end{tcolorbox}

Once the optimal  bid $\bm{W}^\star$ is obtained from Problem \textit{BiD}, the system operator first clears the DAM \eqref{prob:DA} using $\bm{W}^\star$, and then, closer to real time, clears the RTM \eqref{prob:RT} for a particular realization of renewable generation. The expected system cost is denoted by $S^{\text{BiD}}$.

\subsection{Dispatch Benchmarks}

We consider two dispatch  benchmarks for \textit{BiD}:
\subsubsection{Myopic dispatch (MyD)} Each VRES producer $k$ offers the bidding quantity at the expected value of the forecast, i.e.,   ${W}_{k,t}=\mathbb{E}_{\omega \in \Omega}[\widetilde{W}_{k,t,\omega}]$ in \eqref{eq:daw}. Then, the DAM is cleared, followed by the RTM clearing.  The system cost $S^{\text{MyD}}$ is calculated by  $f^{\text{DA}} + \mathbb{E}_{\omega\in \Omega}\left[f_\omega^{\text{RT}}\right]$ following the market clearing problems \eqref{prob:DA} and \eqref{prob:RT}. 

\subsubsection{Stochastic dispatch (StD)}
The system operator co-optimizes the day-ahead schedule and real-time re-dispatch by minimizing the total expected costs across the two markets.  {Note that instead of clearing DAM and RTM separately, the model of \textit{StD} jointly optimized DAM and RTM, which is not directly compatible with the two-settlement market structure.}
\begin{tcolorbox}[standard jigsaw,opacityback=0]
\vspace{-1ex}
\textbf{Problem \textit{StD}: Stochastic dispatch }
\begin{subequations}
\begin{align}
\hspace{-4ex}S^{\text{StD}}:=	{\text{min}} ~&
	f^{\text{DA}}(\Phi^{\text{DA}}) + \mathbb{E}_{\omega\in \Omega}\left[f_\omega^{\text{RT}}(\Phi_\omega^{\text{RT}},\Phi_\omega^{\text{DA}})\right] \\
	\text{s.t.}	~&\Phi_\omega^{\text{RT}}\in \mathcal{X}_\omega^{\text{RT}}(\Phi^{\text{DA}}),~\forall \omega \in \Omega,\\
	&\Phi^{\text{DA}}\in \mathcal{X}^{\text{DA}}(\overline{\bm{W}}),\\
 \text{var:} ~&\Phi^{\text{DA}},~\Phi^{\text{RT}}. \notag
\end{align} 
\end{subequations}
\vspace{-4ex}
\end{tcolorbox}

\textit{System cost comparison:} In terms of the system costs under  \textit{StD}, \textit{BiD} and \textit{MyD}, we have the following comparison.
\begin{prop}
The system costs satisfy  $S^{\text{MyD}}\geq S^{\text{BiD}}\geq S^{\text{StD}}$.
\end{prop}

The reason is that the solution of $\textit{MyD}$ is one feasible solution to Problem \textit{BiD} while the optimal solution to Problem \textit{BiD} is also one feasible solution to Problem \textit{StD}. The two benchmarks serve as the upper bound and lower bound for the proposed bilevel framework.

 {The inequality $S^{\text{MyD}}\geq S^{\text{BiD}}$ is straightforward. We elaborate more on $S^{\text{BiD}}\geq S^{\text{StD}}$. Notably,  Problem \textit{BiD} and Problem \textit{StD} have the same objective but different constraint sets. In  Problem \textit{BiD}, the day-ahead dispatch $\Phi^{\text{DA}}$ is the optimal solution to the deterministic DAM market clearing problem. However, in Problem \textit{StD}, we do not require that  $\Phi^{\text{DA}}$ is the optimal solution to the DAM clearing problem.  If $\Phi^{\text{DA}}$ is feasible in  Problem \textit{BiD}, it means that $\Phi^{\text{DA}}$ satisfies $\Phi^{\text{DA}}\in \mathcal{X}^{\text{DA}}({\bm{W}})$. Then, $\Phi^{\text{DA}}$ will also satisfy $\Phi^{\text{DA}}\in \mathcal{X}^{\text{DA}}(\overline{\bm{W}})$ in Problem \textit{StD}. In other words, the solution space of Problem \textit{BiD} is a subset of Problem \textit{StD}.
 Therefore, Problem \textit{StD} always induces a lower system cost than  Problem \textit{BiD}.}

\section{Solution method}\label{section:solution}

The bilevel Problem \textit{BiD} is non-convex and challenging to solve. The conventional method is to replace the lower-level problem using KKT conditions and transform the problem into a MILP problem\cite{morales2014electricity}. While this method works well on a small-scale system (e.g., IEEE 118-bus system), it does not solve a large-scale system (e.g., NYISO) in a reasonable time \cite{zhao2022uncertainty}. As a solution,  we propose a method based on the strong duality of LP and a McCormick-envelope relaxation.

\subsection{Strong Duality Transformation}
Based on the strong-duality property of LP, we  derive a single-level optimization, which is equivalent to Problem \textit{BiD}.

First, the lower-level LP problem, can equivalently be represented through the set of its KKT conditions, including primary feasibility, dual feasibility, stationary conditions, and complementarity slackness constraints. For LP, the strong-duality condition is equivalent to complementarity slackness constraints \cite{boyd2004convex}. The KKT conditions then take the form: 

(i) Primary feasibility: \eqref{eq:dabalance}-\eqref{eq:darp_b}; 

(ii) Dual feasibility:
\begin{align}
    \tilde{\bm{\lambda}} \geqslant 0,\label{eq:duala}
\end{align}
where $\tilde{\bm{\lambda}}$ includes the set of dual variables $\bm{\lambda}$ except those associated with the power balance constraint \eqref{eq:dabalance}; 

(iii) Stationary conditions, i.e., the first-order conditions with respect to variables $p_{i,t}^{\text{C}},~c_{i,t}^\text{DA}, ~u_{i,t}^{\text{DA}}, ~p_{k,t}^{\text{W}}$, and  $\delta_{n,t}^{\text{DA}}$:
\begin{subequations}\label{eq:stat}
\begin{align}
\frac{\partial\mathcal{L}}{\partial p_{i,t}^{\text{C}}}&=C_{i,t}- \lambda_n^b-\underline{\lambda}_{i,t}^\text{C}+\overline{\lambda}_{i,t}^\text{C}\nonumber\\
&-\overline{\lambda}_{i,t+1}^\text{R}+\underline{\lambda}_{i,t+1}^\text{R}+\overline{\lambda}_{i,t}^\text{R}-\underline{\lambda}_{i,t}^\text{R}
    =0,\label{eq:stata}\\
&\pushright{\forall n \in \mathcal{N},\forall i \in \mathcal{I}_n,\forall t \in \mathcal{T},} \nonumber\\
\frac{\partial\mathcal{L}}{\partial c_{i,t}^\text{DA}}&=1-\lambda_{i,t}^{\text{S}_1}-\lambda_{i,t}^{\text{S}_2}=0,\\
&\pushright{\forall i \in \mathcal{I},\forall t \in \mathcal{T};}\nonumber\\
\frac{\partial\mathcal{L}}{\partial  u_{i,t}^{\text{DA}}}&=C_i^0+\overline{\lambda}_{i,t}^\text{U}-\underline{\lambda}_{i,t}^\text{U}-\overline{P}_i^{\text{C}}\cdot \overline{\lambda}_{i,t}^\text{C}+\underline{P}_i^{\text{C}}\cdot \underline{\lambda}_{i,t}^\text{C} - R_i^\text{U}\cdot \overline{\lambda}_{i,t}^\text{R} \notag\\
&-R_i^\text{D}\cdot \underline{\lambda}_{i,t+1}^\text{R} 
+C_i^{\text{SU}}\cdot \lambda_{i,t}^{\text{S}_1}-C_i^{\text{SU}}\cdot \lambda_{i,t+1}^{\text{S}_1}=0,\\
&\pushright{\forall i \in \mathcal{I},~\forall t \in \mathcal{T},}\nonumber\\
\frac{\partial\mathcal{L}}{\partial  p_{k,t}^{\text{W}}}&=- \lambda_{n,t}^b-\underline{\lambda}_{k,t}^\text{W}+\overline{\lambda}_{k,t}^\text{W}=0,\label{eq:staw}\\
&\pushright{\forall n \in \mathcal{N},\forall k \in \mathcal{K}_n,\forall t \in \mathcal{T},}\nonumber\\
\frac{\partial\mathcal{L}}{\partial \delta_{n,t}^{\text{DA}}}&=\sum_{m:(n,m)\in\Lambda}\hspace{-3ex}\frac{-\lambda_{n,t}^b\!-\!\underline{\lambda}_{n,m,t}\!+\!\overline{\lambda}_{n,m,t}}{x_{n,m}}\nonumber\\
&\quad\quad-\sum_{m:(m,n)\in\Lambda}\hspace{-2ex}\frac{-\lambda_{m,t}^b\!-\!\underline{\lambda}_{mn,t}\!+\!\overline{\lambda}_{mn,t}}{x_{mn}}=0,\label{eq:statz}\\
&\pushright{\forall n \in \mathcal{N},\forall t \in \mathcal{T};}\nonumber
\end{align}
\end{subequations}
(iv) Strong-duality condition:
    \begin{align}
        f^{\text{DA}}(\Phi^{\text{DA}}) = g^{\text{DA}}(\bm{\lambda}), \label{eq:strongdual}
    \end{align}
    where the dual objective function is defined by 
        \begin{align}
     g^{\text{DA}}(\bm{\lambda}):=&\sum_{n\in \mathcal{N}}\!\sum_{t\in \mathcal{T}} \lambda_{n,t}^b\cdot L_{n,t}- \!\!\sum_{n,m\in \mathcal{N}} (\underline{\lambda}_{n,m,t}+\overline{\lambda}_{n,m,t})\cdot \overline{F}_{n,m}\notag\\
     &-\sum_{i\in \mathcal{I}}\sum_{t\in \mathcal{T}} \overline{\lambda}_{i,t}^{\text{U}}-\sum_{k\in \mathcal{K}} \sum_{t\in \mathcal{T}} \overline{\lambda}_{k,t}^\text{W} \cdot W_{k,t}.\label{eq:dualobj}
    \end{align}

Then,  equivalent to Problem \textit{BiD}, the single-level optimization Problem \textit{BiD-S}  takes the following form: 
\begin{tcolorbox}[standard jigsaw,opacityback=0]
\hspace{-2ex}\textbf{Problem \textit{BiD-S}: Equivalence with strong duality}
\begin{subequations} \label{prob:bilevel_new}	
	\begin{align}
		\min  ~&
		f^{\text{DA}}(\Phi^{\text{DA}}) + \mathbb{E}_{\omega\in \Omega}\left[f_\omega^{\text{RT}}(\Phi_\omega^{\text{RT}},\Phi_\omega^{\text{DA}})\right] \notag \\	
	\text{s.t.} ~& \Phi_\omega^{\text{RT}}\in \mathcal{X}_\omega^{\text{RT}}(\Phi^{\text{DA}}),~\forall \omega \in \Omega,\notag\\
		&\Phi^{\text{DA}}\in \mathcal{X}^{\text{DA}}(\bm{W}),\notag\\
  &\text{Equations}~\eqref{eq:duala}, ~\eqref{eq:stat}, ~\eqref{eq:strongdual},\notag\\
  \text{var}: ~&\Phi^{\text{RT}}, \Phi^{\text{DA}}, \bm{W}, \bm{\lambda}. \notag
	\end{align}

\end{subequations}
\end{tcolorbox}
However, Problem \textit{BiD-S} has the  only bilinear term $\overline{\lambda}_{k,t}^W \cdot W_{k,t}$ in the strong-duality constraint \eqref{eq:strongdual}. Next, we  adopt the McCormick envelope to relax this bilinear item.

\subsection{McCormick-Envelope Relaxation}
We will first formulate the McCormick-envelope relaxation, and then explain how we choose the bounds for the envelope. 

\subsubsection{Convex relaxation} We let $z_{k,t}=\overline{\lambda}_{k,t}^\text{W} \cdot W_{k,t}$ in \eqref{eq:dualobj} and change $g^{\text{DA}}(\bm{\lambda})$ into $g^{\text{DA}}(\bm{\lambda},\bm{z})$, which transforms \eqref{eq:strongdual}  into   
\begin{align}
        f^{\text{DA}}(\Phi^{\text{DA}}) =  g^{\text{DA}}(\bm{\lambda},\bm{z}). \label{eq:strongdualn}
    \end{align}

Given some variable bounds $\alpha_{k,t}^\lambda\leqslant \overline{\lambda}_{k,t}^\text{W} \leqslant \beta_{k,t}^\lambda$ and $\alpha_{k,t}^\text{W}\leqslant  W_{k,t} \leqslant \beta_{k,t}^\text{W}$, the McCormick envelope gives the following  convex relaxation for  $z_{k,t}$ \cite{mccormick1976computability}.
\begin{subequations}\label{eq:mc_all}
\begin{align}
&z_{k,t} \geqslant \alpha_{k,t}^\lambda \cdot W_{k,t}+ \alpha_{k,t}^\text{W} \cdot \overline{\lambda}_{k,t}^\text{W}  -\alpha_{k,t}^\lambda \alpha_{k,t}^\text{W},\label{eq:mca}\\
&z_{k,t} \geqslant \beta_{k,t}^\lambda \cdot W_{k,t}+ \beta_{k,t}^\text{W}\cdot \overline{\lambda}_{k,t}^\text{W}  -\beta_{k,t}^\lambda \beta_{k,t}^\text{W},\\
&z_{k,t} \leqslant \beta_{k,t}^\lambda \cdot W_{k,t}+ \alpha_{k,t}^\text{W}\cdot \overline{\lambda}_{k,t}^\text{W}  -\beta_{k,t}^\lambda \alpha_{k,t}^\text{W},\\
&z_{k,t} \leqslant \alpha_{k,t}^\lambda \cdot W_{k,t}+\beta_{k,t}^\text{W}\cdot \overline{\lambda}_{k,t}^\text{W}  -\alpha_{k,t}^\lambda \beta_{k,t}^\text{W}, \label{eq:mcz}\\
&\pushright{\forall k \in \mathcal{K},\forall t \in \mathcal{T},}\nonumber
\end{align}
\end{subequations}
This leads to the following relaxed LP counterpart Problem \textit{BiD-McCormick} of Problem \textit{BiD-S}:
\vspace{-1ex}
\begin{tcolorbox}[standard jigsaw,opacityback=0]
\textbf{
\hspace{-1ex}Problem \textit{BiD-McCormick}: Relaxed problem}
\vspace{-0.8ex}
\begin{subequations}	
	\begin{align}
		\min  ~&
		f^{\text{DA}}(\Phi^{\text{DA}}) + \mathbb{E}_{\omega\in \Omega}\left[f_\omega^{\text{RT}}(\Phi_\omega^{\text{RT}},\Phi_\omega^{\text{DA}})\right] \notag \\	
	\text{s.t.} ~& \Phi_\omega^{\text{RT}}\in \mathcal{X}_\omega^{\text{RT}}(\Phi^{\text{DA}}),~\forall \omega \in \Omega,\notag\\
		&\Phi^{\text{DA}}\in \mathcal{X}^{\text{DA}}(\bm{W}),\notag\\
  &\text{Equations}~\eqref{eq:duala}, ~\eqref{eq:stat}, ~\eqref{eq:strongdualn},~\eqref{eq:mc_all}\notag\\
  \text{var}: ~&\Phi^{\text{DA}},\Phi^{\text{RT}}, \bm{W}, \bm{\lambda},\bm{z}. \notag
	\end{align}
\end{subequations}
\end{tcolorbox}

\subsubsection{Bounds choice} It is important to choose proper lower and upper bounds for $W_{k,t}$ and  $\overline{\lambda}_{k,t}^\text{W}$.   In numerical studies, for $W_{k,t}$,  we choose the bounds $\alpha_{k,t}^\text{W}=0$ as $ 0\leqslant W_{k,t}$. We set
\vspace{-1ex}
\begin{align}
  \beta_{k,t}^\text{W}=\gamma \cdot \mathbb{E}_{\omega}[\widetilde{W}_{k,t,\omega}],~\forall k\in \mathcal{K},\forall t\in \mathcal{T},\label{eq:gamma}
\end{align}
where we adjust the parameter $\gamma$ for the upper bounds around the mean value of VRES forecast to test the performance.
For $\overline{\lambda}_{k,t}^\text{W}$, we let $\alpha_{k,t}^\lambda=0$  as $0\leqslant \overline{\lambda}_{k,t}^\text{W}$. We set $\beta_{k,t}^\lambda=\overline{\lambda}^\text{W}_{k,t}(\bm{0})$, where $\overline{\lambda}^\text{W}_{k,t}(\bm{0})$ denotes the dual solution of $\overline{\lambda}^\text{W}_{k,t}$ in the DAM schedule when all VRES producers bid zero quantity. Based on \eqref{eq:staw}, when $\overline{\lambda}_{k,t}^\text{W}>0$, we have $\overline{\lambda}_{k,t}^\text{W}=\lambda_{n,t}^b, \forall k \in \mathcal{K}_n$. Since VRES have zero marginal cost in the DAM, more renewable energy at one bus indicates lower locational marginal prices. Thus, the dual value $\overline{\lambda}^\text{W}_{k,t}(\bm{0})$ can roughly indicate the upper bound for $\overline{\lambda}^\text{W}_{k,t}$, which is always  feasible in simulations.

\subsubsection{System cost $S^{\text{BiD}}$ calculation} After obtaining the quantity solution $\bm{W}^\star$ from Problem \textit{BiD-McCormick}, the system operator will first clear the DAM \eqref{prob:DA} and obtain the DAM cost $f^{\text{DA}}$ given $\bm{W}^\star$, and then sequentially clear the RTM \eqref{prob:RT} and obtain the RTM cost $f_\omega^{\text{RT}}$, based on which we calculate the system cost $S^{\text{BiD}}$ by  $f^{\text{DA}} + \mathbb{E}_{\omega\in \Omega}\left[f_\omega^{\text{RT}}\right]$.\footnote{ {Although we solve the whole problem by linear relaxation, what we aim to obtain is  the optimal bidding quantity of VRES. Given the optimal bidding quantity, we re-run the DAM  (lower-level problem) and RTM models to clear the market. Therefore, the market-clearing solutions in the lower level and upper level, as resulting from these market simulations, are always feasible.}} We use the same scenario set to calculate $S^{\text{BiD}}$ as in Problem \textit{BiD-McCormick}. Note that we are not using the optimal objective value in Problem \textit{BiD-McCormick} as the system cost for $S^{\text{BiD}}$.

\subsubsection{ {Approximation error of McCormick envelope}} 
 {Note that for Problem \textit{BiD}, the KKT-condition-based reformulation is equivalent to the original problem. However,  we may not be able to numerically solve the problem for a large-scale system. e.g., the  NYISO system that we use in Section \ref{section:simulation}. Therefore, we cannot use this result as a benchmark for our McCormick-envelope relaxation.\footnote{For a small-scale system like IEEE 118-bus system, as shown in the previous conference paper \cite{zhao2022uncertainty}, the gap is no greater than 2\% between the McCormick-envelope-based method and accurate KKT-condition-based method.} Besides, we  cannot theoretically characterize a bound for the relaxation error, which is a topic for future research.}

 {However, we have shown that \textit{StD} provides a natural lower bound for \textit{BiD} in terms of the expected system cost. Hence the approximation error (in system cost) of the McCormick-envelope relaxation for the bilevel model is upper-bounded by the difference between Problem  \textit{BiD-McCormick} and Problem \textit{StD}. We use such an upper bound of the error to evaluate the performance of Problem  \textit{BiD-McCormick}, which is usually less than 6\% in the simulations in Section \ref{section:simulation}.}

\subsubsection{ {Application to other problems}}  {The proposed McCormick-envelope relaxation method can be  applied to other bilevel
optimization problems. First, as long as the lower-level problem is a linear program, we can use the strong-duality condition to obtain the equivalent transformations. Second, the strong-duality condition brings bilinear terms. It is important to choose proper upper bounds and lower bounds for the McCormick envelope. We select upper and lower bounds based on our insights into the VRES scheduling problem. When we apply this method to other applications, proper upper and lower bounds need to be selected considering their specific physical meanings. Lastly, this method can work well on a specific problem if it has good accuracy performance, which is the case for our application as shown later in Section \ref{section:simulation}.  However, if the accuracy performance is poor for a specific problem, we can try multi-segment McCormick envelopes, which introduce piece-wise linear constraints instead of linear constraints. In this case, the tradeoff between accuracy and complexity needs to be evaluated.}

In the next section, we will show that even if the selected lower and upper bounds for $ W_{k,t} $ and  $\overline{\lambda}_{k,t}^\text{W}$ are in wide ranges, they can still achieve system costs that are much lower than \textit{MyD} and close to \textit{StD} as shown in Section \ref{sec:sim:accuracy}. Besides, the relaxed Problem \textit{BiD-McCormick} is an LP problem that can be efficiently solved even on large-scale systems.

\section{Numerical studies}\label{section:simulation}
We test our \textit{BiD} framework on the NYISO system \cite{greene2022}. We will first show a good performance of the proposed algorithm based on strong-duality conditions and McCormick envelopes. Then, we demonstrate the economic benefits of \textit{BiD} compared with the baseline \textit{MyD}, e.g.,  reducing system costs, improving robustness under out-of-sample tests for uncertainty scenarios, reducing  volatility of  market prices and revenues, and preventing revenue loss for VRES producers.
\subsection{Modeling Setup}

\textit{NYISO system:} The tested  NYISO system has 1576 buses, 2359 transmission lines, 1564 loads, 345 conventional generation units,  and 27 wind farms \cite{greene2022}. We use the wind and load data on August 2, 2019. For illustration, we adopt the operation window from 7:00 am to 10:00 am. We generate 20 joint scenarios for the probability distribution of wind and load forecast using PGscen \cite{carmona2022joint}. The simulations on 24-hour operations and a large number of scenarios are computationally challenging.  We give one example of 24-hour schedule results based on 4-hour rolling horizons in Appendix A. We present out-of-sample tests for 20 scenarios in Section \ref{sec:sim:outofsample}.

\textit{Wind-energy generators:} We have 27 wind farms in NYISO with 20 scenarios representing uncertainty in their generations. Since the original capacity of wind farms is small, i.e., it only meets 4.4\% of the total demand, we will gradually increase the capacity of wind farms to achieve penetration levels of 10\%, 40\%, and 70\%, relative to system demand on average. To accommodate more renewable energy, we also increase transmission-line capacities by factors $\times$ 2 and $\times$ 4.

\textit{Flexibility of conventional generators:} We classify conventional generators into fast-startup and slow-startup technologies: those using steam turbines and combined-cycle systems are regarded as slow-startup generators and the rest, including gas turbines, internal combustion, and hydro, are fast-startup generators. In the NYISO system, there are 582 conventional units, among which there are 126 slow-startup units whose capacity makes up 26\% of the total generation capacity. In the interest of experiments, we also consider varying flexibility for the conventional generation units. 

\textit{Additional Notation:} We use notation \textit{xR} and \textit{yL} to indicate the $x\%$ percentage increase of VRES power capacity and multiplicative factor $y$ for the scaled transmission line capacity, respectively. For example, {\it 70R-2L} denotes 70\% wind power penetration with line capacities scaled by a factor of 2. In terms of varying generation flexibility, we use \textit{mFlx} to denote the current state of medium generator flexibility, \textit{lFlx} to denote the low flexibility case with all the generators as slow start-up units, and \textit{hFlx} to denote the high flexibility case where all the generators are fast-startup units.

We simulate on a MacBook Pro (2020) with a 2.3 GHz Quad-Core Intel Core i7 processor. We use the solver Gurobi to solve the LPs  in Julia/JuMP.

\subsection{Performance of \textit{BiD-McCormick}}\label{sec:sim:accuracy}


We show that the proposed relaxation technique for \textit{BiD} can achieve good performance in accuracy and scalability.

We simulate a case of high levels of VRES at \textit{40R-1L-mFlx}. Figures \ref{fig:bounds}a and \ref{fig:bounds}b show the expected system costs and day-ahead scheduled  VRES amounts, respectively, where we adjust the upper bound parameter $\gamma$ in \eqref{eq:gamma}. Different curves represent \textit{StD},  \textit{BiD}, and \textit{MyD}, respectively. Table \ref{tab:cpu_nyiso} shows the computation time for different scenarios and operation times.

\textit{Accuracy:} As we increase $\gamma$ in a wide range from 0.2  to 1.4, in Figure \ref{fig:bounds}a, the expected system cost under \textit{BiD} (blue curve) is very close to \textit{StD} (green curve)  within a gap of $6\%$, but lower than \textit{MyD} (red curve) by over 20\%. This shows that \textit{BiD} can achieve a performance close to the least-cost benchmark \textit{StD}, which is also robust under the bound choice of  $\gamma$. As shown in Figure \ref{fig:bounds}b, the day-ahead scheduled VRES quantities under \textit{StD} and \textit{BiD} are more conservative than \textit{MyD}  to reduce the re-dispatch cost in the RTM.  {We also discuss the impact of the upper bound of dual variables in the McCormick envelope in Appendix \ref{app:dual}.}

\textit{Scalability:} Our preliminary results \cite{zhao2022uncertainty} have shown that the conventional method of reformulating a MILP problem based on KKT conditions cannot solve the large-scale NYISO system within two hours, even for an economic dispatch problems with only 5 wind scenarios. In our extended UC problem, Table \ref{tab:cpu_nyiso} shows that \textit{BiD} can still compute the result within 20 minutes for 20 scenarios and 4 operation hours.\footnote{Different values of $\gamma\in [0,1.6]$ do not change running time significantly.} However, we acknowledge that as the number of scenarios and operation hours increases, both \textit{BiD} and \textit{StD} may face computational challenges.

 {To further facilitate the computation of large-scale stochastic linear programs, some decomposition techniques can be potentially applied to decompose across scenarios and hours so that the problem can be solved more efficiently in a distributed manner. These methods include progressive hedging algorithms to decompose scenarios \cite{ordoudis2015stochastic,gu2016stochastic}, the lagrangian relaxation method for temporal decomposition \cite{kim2018temporal}, and the 
alternating direction method of multipliers (ADMM) to decompose network and temporal constraints \cite{feizollahi2015large}. In addition to conventional optimization techniques, machine learning methods are now also gaining popularity in efficiently solving large-scale optimization problems, e.g., using neural networks to solve DC power flow problems \cite{pan2021deepopf} or directly solve bilevel programs \cite{molan2023using}. These machine learning techniques can be explored in the future to solve the bilevel and linear programs more efficiently in power systems.}

\begin{figure}[t]
	\centering
	\hspace{-2ex}
	\subfigure[]{
		\raisebox{-2mm}{\includegraphics[width=1.72in]{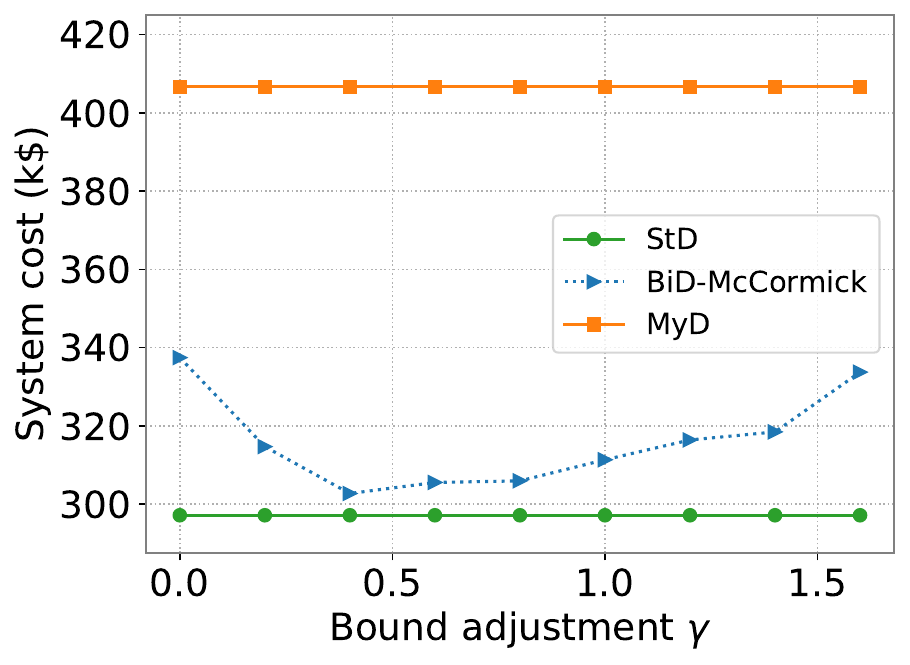}}}
	\hspace{-1ex}
	\subfigure[]{
		\raisebox{-2mm}{\includegraphics[width=1.72in]{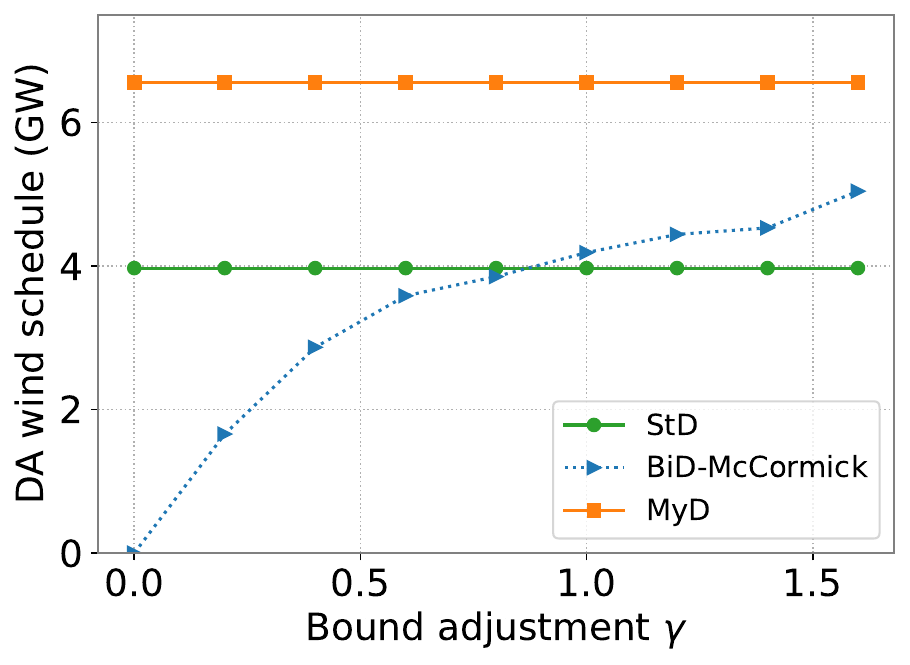}}}
	\vspace{-3mm}
	\caption{ (a) \small Expected system cost; (b) Day-ahead scheduled total VRES quantity. Both are functions of bound adjustment $\gamma$.}
	\label{fig:bounds}
\end{figure} 

\begin{table}[bt!]
\caption{NYISO: Computation time (minutes) for a varying number of uncertainty scenarios and operation hours.}
\vspace{-2ex}
\label{tab:cpu_nyiso}
\begin{center}
\begin{tabular}{lcccccccc}
\toprule
\# Scenarios \# Hours & 5S2H & 10S2H  & 20S2H  & 20S4H & 20S6H\\
\midrule
\textit{MyD}      & 0.35  &  0.35  &  0.35  &  0.53 &1.0  \\
\textit{BiD} & 2.0  & 3.1   & 7.4  & 19.4 &48.9 \\
\textit{StD}   & 0.76 &  1.2  &  3.0  &  7.3 &16.3\\
\bottomrule
\end{tabular}
\end{center}
\vspace{-3ex}
\end{table}

\begin{figure*}[t]
	\centering
	\hspace{-2ex}
	\subfigure[]{
		\raisebox{-2mm}{\includegraphics[width=2.45in]{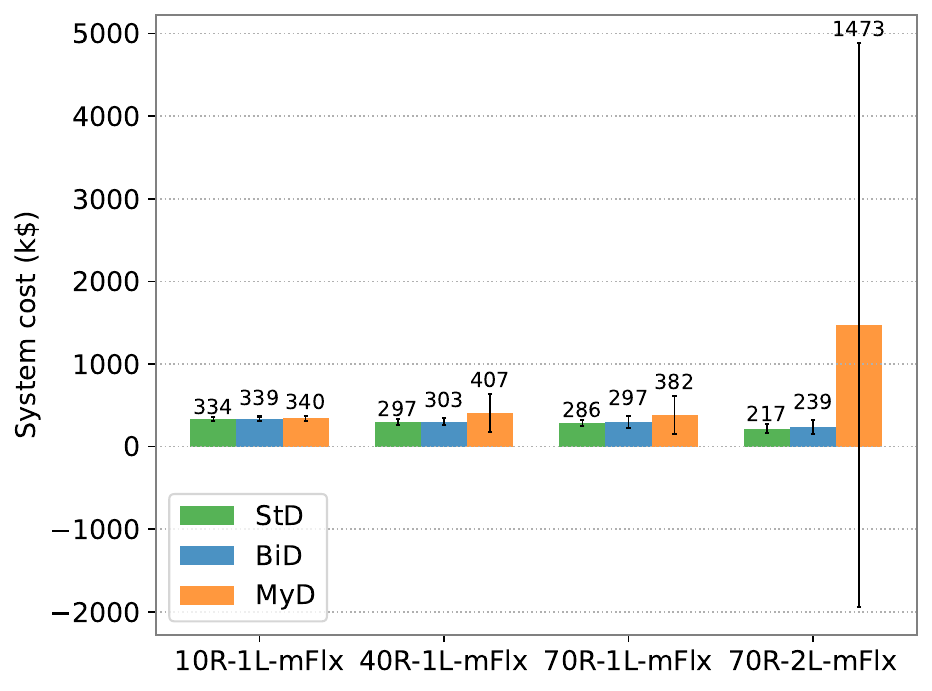}}}
	\hspace{-1ex}
	\subfigure[]{
		\raisebox{-2mm}{\includegraphics[width=2.45in]{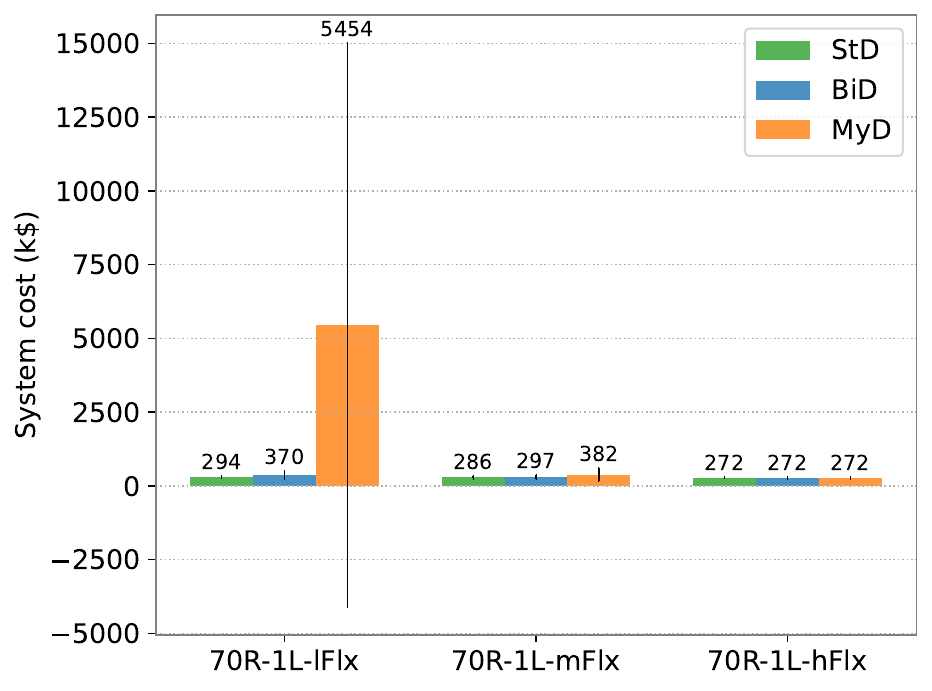}}}
	\vspace{-2mm}
	\caption{(a) \small Under \textit{mFlx}: expected system cost under different wind-energy amounts (-\textit{R}) and transmission line capacities (-\textit{L}); (b) \small Under \textit{70R-1L}: expected system cost under different generator flexibility levels (-\textit{Flx}).}
	\label{fig:sys_cost}
 \vspace{-1ex}
\end{figure*} 

\begin{table}[bt!]
\caption{\textit{BiD} vs \textit{MyD}: System cost reduction.}
\vspace{-2ex}
\label{tab:cost_reduction}
\begin{center}
\begin{tabular}{lcccc}
\toprule

Under \textit{mFlx} & System cost reduction \\
\midrule
\textit{10R-1L}     & 0.3\%  \\
\textit{40R-1L} & 25\%  \\
\textit{70R-1L} & 23\% \\
\textit{70R-2L}   & 84\% \\
\bottomrule
\toprule
Under \textit{70R-1L} & System cost reduction \\
\midrule
\textit{lFlx}     & 93\%  \\
\textit{mFlx} & 23\%  \\
\textit{hFlx} & -0.4\% \\

\bottomrule
\end{tabular}
\end{center}
\vspace{-5ex}
\end{table}

\begin{figure*}[t]
	\centering
	\hspace{-2ex}
  	\subfigure[]{
      		\raisebox{-2mm}{\includegraphics[width=2.7in]{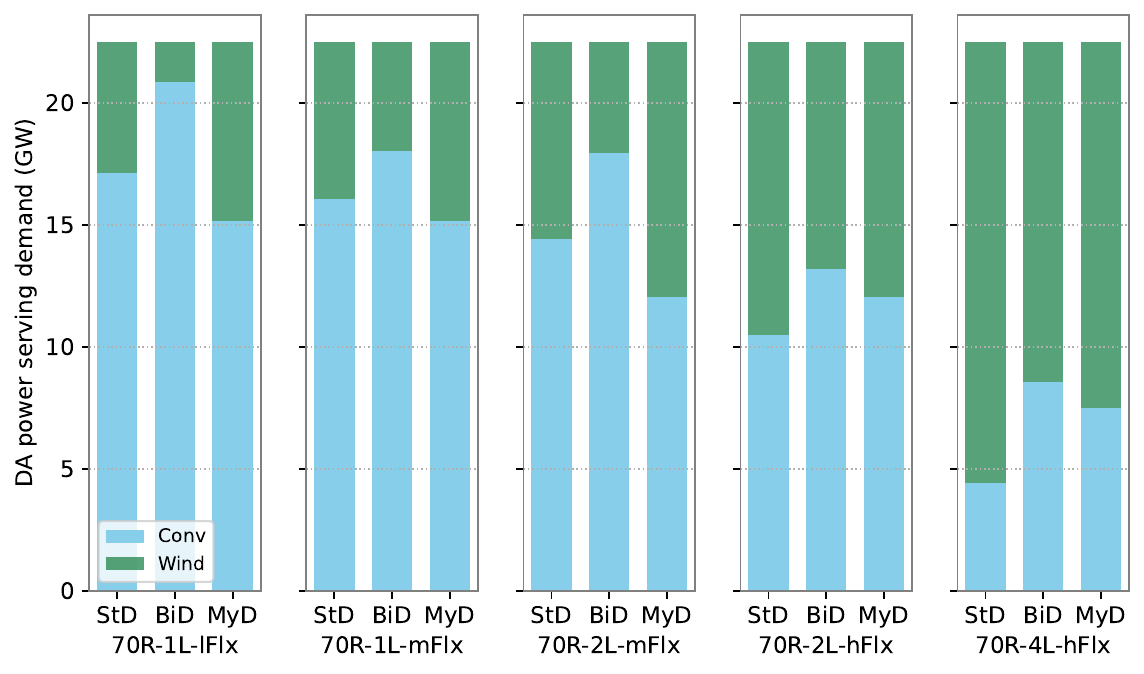}}}
          	\subfigure[]{
      		\raisebox{-2mm}{\includegraphics[width=2.7in]{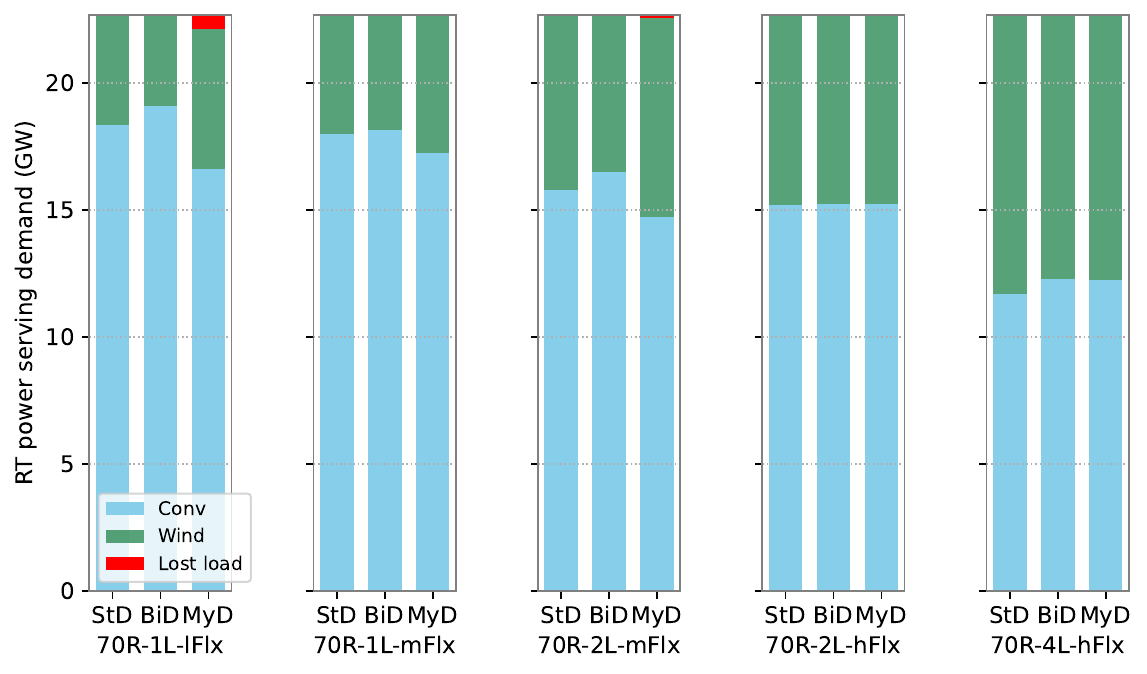}}}
	\vspace{-2ex}
	\caption{\small (a) DAM schedule to serve demand  from different sources. (b) Actual power to serve demand in real time from different sources.}
	\label{fig:sys_power}
 \vspace{-1ex}
\end{figure*}

\subsection{BiD Performance: System Cost Reduction}
We show that the \textit{BiD} framework can substantially reduce the expected system cost compared with \textit{MyD}. However, this benefit is affected by the factors of VRES amounts and generator flexibility levels. Overall, a higher VRES penetration level and a lower flexibility level of generators will increase the performance gap between \textit{BiD} and \textit{MyD}. Furthermore,  increasing transmission-line capacities may result in a much higher system cost under \textit{MyD}  (while a lower cost under \textit{BiD}) because of the lack of flexible generators in real-time to deal with uncertainty.

In Figure \ref{fig:sys_cost}, we demonstrate the expected system costs under different parameter settings. The error bars show the standard deviation (std) of the results.  In Figure \ref{fig:sys_cost}a, we fix the flexibility level of generators at \textit{mFlx}. We show the expected system costs under \textit{StD}, \textit{BiD}, and \textit{MyD}, respectively, considering different VRES amounts and transmission-line capacities.  In Figure \ref{fig:sys_cost}b, we fix the VRES amounts and transmission-line capacity at \textit{70R-1L} and vary different flexibility levels of generators. In Table \ref{tab:cost_reduction},  we show the values of system cost reduction when comparing \textit{BiD} with \textit{MyD}. 

In Figure \ref{fig:sys_power}, under different parameters, we show the load-serving power amounts (average over hours and scenarios) from different sources. Figure \ref{fig:sys_power}a shows 
DAM scheduled quantities of conventional generators and wind energy. Figure \ref{fig:sys_power}b shows actual power amounts from conventional generators, wind energy,  and load shedding, to serve real-time load.  We have the following observations.

\begin{figure*}[t]
	\centering
	\hspace{-2ex}
	\subfigure[]{
		\raisebox{-2mm}{\includegraphics[width=2.35in]{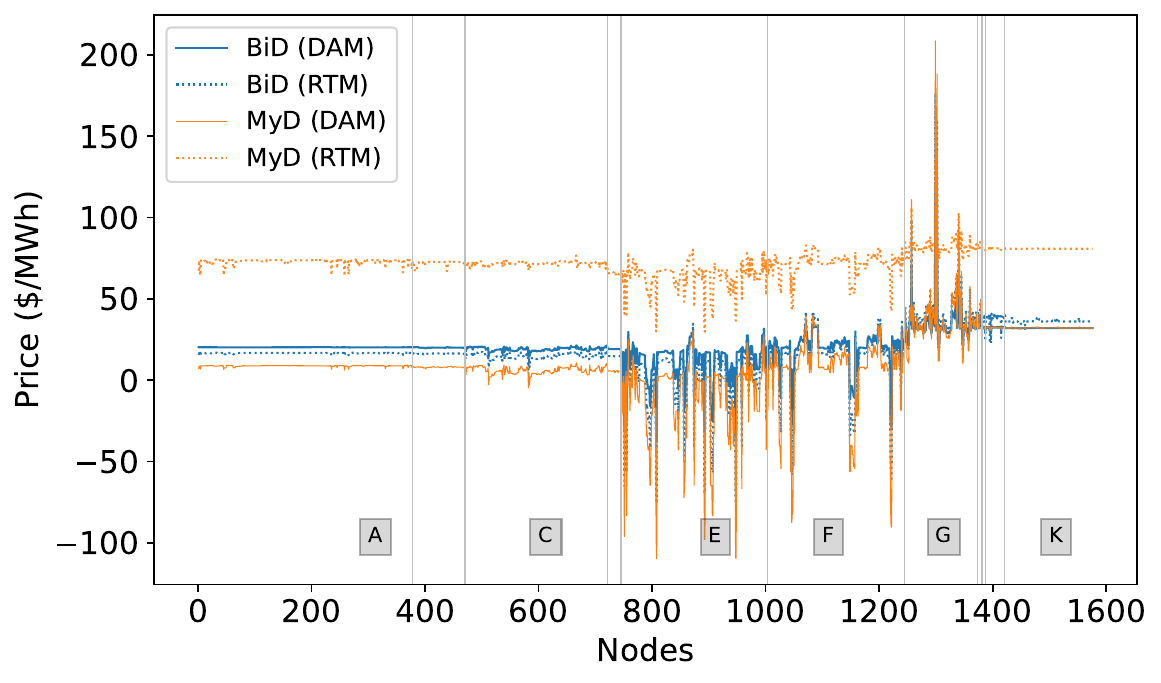}}}
	\hspace{-2ex}
	\subfigure[]{
		\raisebox{-2mm}{\includegraphics[width=2.35in]{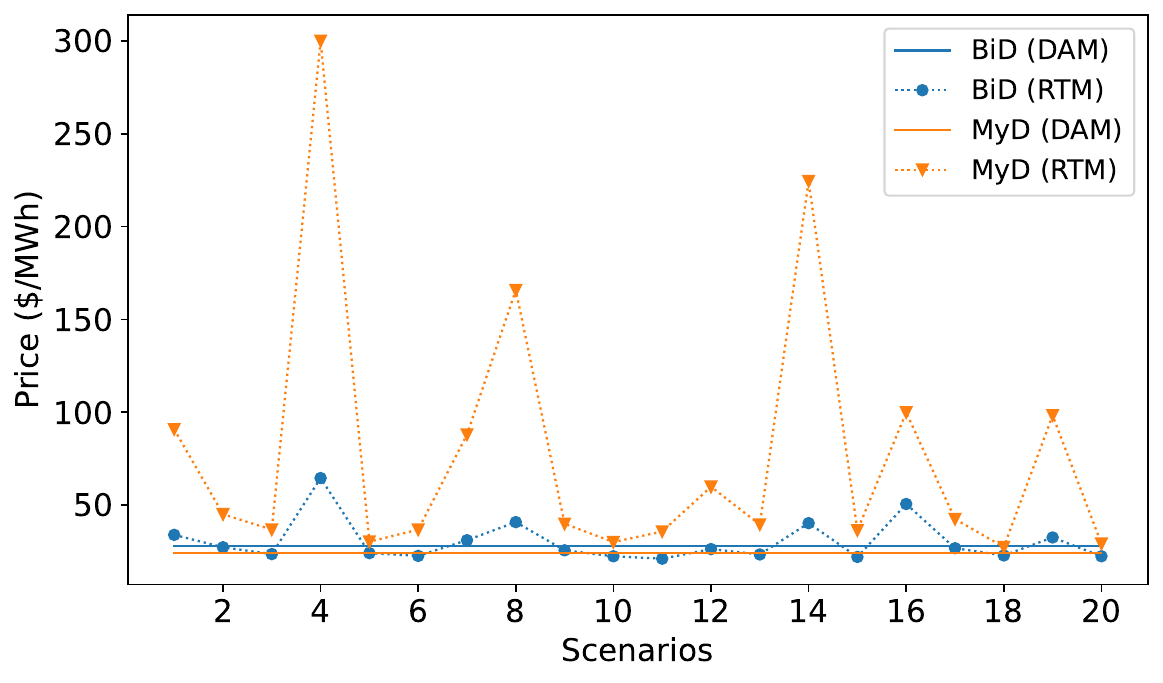}}}
  \hspace{-2ex}
  	\subfigure[]{
		\raisebox{-1mm}{\includegraphics[width=2.35in]{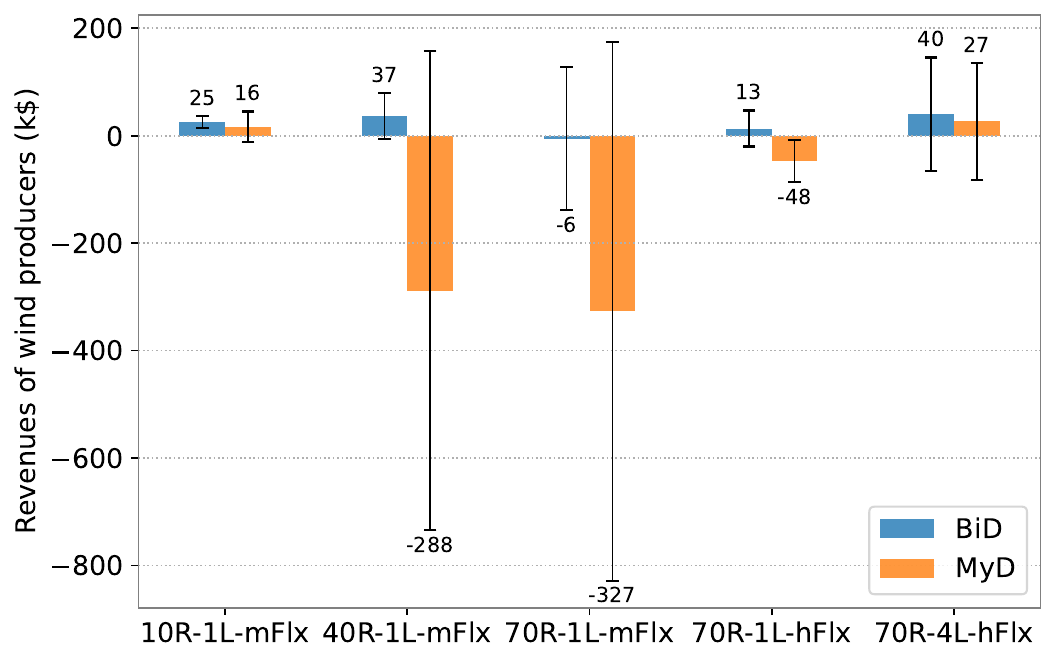}}}
  \vspace{-1ex}
	\caption{ \textit{BiD} vs \textit{MyD}: (a) \small Hourly DAM and RTM prices (average in scenarios)  under different nodes, where A-K show  different regions of NYISO (with parameter setting \textit{40R-1L-mFlx}); (b) \small Hourly DAM and RTM prices (average in nodes)  under different scenarios (with parameter setting \textit{40R-1L-mFlx}). (c) Wind producers' revenues (total in DAM and RTM) under different parameter settings.}
	\label{fig:priceprofitcomp}
 \vspace{-2ex}
\end{figure*}

\textit{System cost reduction:  \textit{BiD} can substantially reduce the expected system cost and its volatility compared with \textit{MyD}, particularly under high VRES generation amounts and limited flexibility of conventional generators.}   \textit{MyD}  cannot account for real-time re-dispatch costs, which prevents it from achieving market efficiency under high VRER amounts and lower flexibility levels of generators. In particular, under \textit{70R-2L-mFlx} (Figure \ref{fig:sys_cost}a) or  \textit{70R-1L-lFlx} (Figure \ref{fig:sys_cost}b), the system cost can be very high under \textit{MyD} due to the lost load in real-time as shown in Figure \ref{fig:sys_power}b. 

\textit{Impact of VRES penetration levels: A higher VRES penetration level will reduce the system cost under \textit{BiD} and \textit{StD}, but can increase the system cost under \textit{MyD} if the system lacks flexible generators.} As shown in Figure \ref{fig:sys_cost}a, the system cost under \textit{BiD} and \textit{StD} always decreases as the VRES amount or the transmission-line capacity increases. However, the system cost under \textit{MyD} does not. For instance, when we increase the transmission line capacity from \textit{1L} to \textit{2L} in Figure \ref{fig:sys_cost}a, the system cost rises significantly under \textit{MyD}. The reason is that the increased line capacity allows more VRES to be dispatched in the DAM and  keeps more slow-startup generators  offline in  the DAM. Then, it leads to load shedding (Figure \ref{fig:sys_power}b: \textit{70R-2L-mFlx}) in real time due to the lack of flexible resources.  Furthermore, note that although the wind-energy generation amount makes up 70\% of the total demand, the transmission limit will lead to the curtailment of renewable energy in RTM. For example, under the case of \textit{70R-2L-hFlx}, the maximum amount serving the demand is about 50\%  in DAM (Figure \ref{fig:sys_power}a) and 30\%  in RTM (Figure \ref{fig:sys_power}b).

\textit{Impact of generator flexibility: Increasing the flexibility level of generators (assuming unchanged generation costs) will reduce the expected system cost for each dispatch framework.  \textit{MyD} can achieve market efficiency close to \textit{BiD} and \textit{StD}  if there is high flexibility.} As shown in Figure \ref{fig:sys_cost}b, for each dispatch framework, the increasing flexibility levels reduce the system cost. In fact, under \textit{hFlx},  \textit{MyD}, \textit{BiD}, and \textit{StD}  will induce the same system cost (Figure \ref{fig:sys_cost}b: \textit{hFlx}) and similar real-time energy-source supply mix (Figure \ref{fig:sys_power}b: \textit{70R-2L-hFlx}), despite the different DAM schedule results (Figure \ref{fig:sys_power}a:\textit{70R-2L-hFlx}).  Under the current flexibility level \textit{mFlx}, i.e., 126 slow-startup units among 582 conventional generators, \textit{MyD} does not work well under high penetration levels of VRES.

\begin{figure}[t]
	\centering
	\hspace{-3ex}
\includegraphics[width=2.2in]{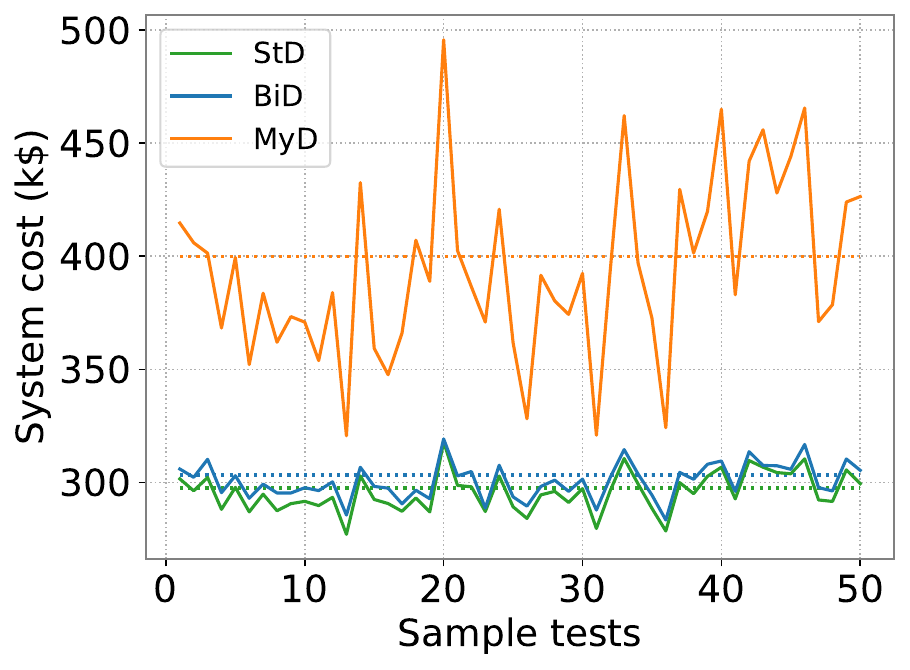}
\vspace{-1ex}
	\caption{Expected system costs under different samples where dash lines are for the initial scenario set.}
	\label{fig:sample}
 \vspace{-3ex}
\end{figure} 

\subsection{Out-of-Sample Test} \label{sec:sim:outofsample}
 
 {Our work  optimizes the DAM bidding strategies of VRES based on the given probability distributions of VRES generators. Different probability distributions will lead to different solutions. The uncertainty of VRES is initially modeled by continuous probability distributions.\footnote{The continuous probability distributions are modeled in PGscen \cite{dataarpae}.} The reason to do the out-of-sample test is that we are using discrete scenarios to approximate the initial continuous probability distributions. The newly generated scenarios for out-of-sample testing are sampled from the same continuous probability distribution as the original scenarios based on the Monte Carlo method.}

We show that \textit{BiD} achieves more robust results than \textit{MyD} in out-of-sample tests. Recall that Figure \ref{fig:sys_cost} shows results based on the initial 20 scenarios of wind power and load forecast. Now, we will test the day-ahead schedule obtained  under the initial  scenario set on other scenario sets. 

In the out-of-sample test, we still use PGscen to generate different scenario sets of wind energy and load. Note that PGscen  uses the Monte Carlo method to generate scenarios.
For each test,  we randomly generate 20 scenarios. Given the day-ahead schedule under the initial scenario set in Section \ref{section:simulation},  we compute the re-dispatch result under the newly generated scenario set and calculate the total system cost. We run 50 tests in total and focus on the setting of  \textit{40R-1L-mFlx}.

In Figure \ref{fig:sample}, we show the system costs under different test samples for \textit{MyD} (red solid curve), \textit{BiD} (blue solid curve), and \textit{StD} (green solid curve), respectively. The dashed flat curves show the system cost under the original scenario set in Section \ref{section:simulation}.  The std values (k\$) of system costs  under \textit{MyD}, \textit{BiD}, and \textit{StD}  are 39.3, 7.9, and 8.6, respectively. Figure \ref{fig:sample} shows that \textit{BiD} and \textit{StD}  can achieve more robust results in terms of system costs under different scenario samples, compared with \textit{MyD}. \textit{MyD} schedules more VRES in DAM and brings more uncertainty.  {In addition, we discuss the impact of the number of scenarios on the out-of-sample test in Appendix \ref{app:scenario}.}

\subsection{BiD Performance: Prices and Revenues}

We demonstrate that \textit{BiD} is effective in reducing the volatility of various market factors when compared to \textit{MyD}. By bringing more conventional generators online  in the DAM, \textit{BiD}  can decrease the volatility of market prices and wind producers' revenues across scenarios.\footnote{We also show the results for conventional generators' revenues and RTM re-dispatch amounts in Appendix \ref{app:volitility}.} Additionally, \textit{BiD} can help to prevent negative revenues for VRES.

 In Figures \ref{fig:priceprofitcomp}a and b, we set the parameters at \textit{40R-1L-mFlx}. We show the market prices at different nodes (average in scenarios) and different scenarios (average in nodes), respectively. The solid and dotted curves show the DAM and RTM prices, respectively. The blue curves show the \textit{BiD} result and the orange ones are for \textit{MyD}.  In Figures \ref{fig:priceprofitcomp}c, we show the revenues  for wind producers under different parameter settings.

\textit{Volatility reduction: \textit{BiD} can reduce the volatility of market prices and  VRES producers' revenues.}  As shown in Figures \ref{fig:priceprofitcomp}a and  \ref{fig:priceprofitcomp}b,  the market prices under \textit{BiD} exhibit less volatility across nodes and scenarios. Moreover, under \textit{BiD}, the RTM prices fluctuate around the DAM prices, whereas \textit{MyD} generates higher RTM prices than DAM in almost all scenarios and nodes. In Figure \ref{fig:priceprofitcomp}c, under high VRES amounts and limited generator flexibility (i.e., \textit{40R-1L-mFlx} and \textit{70R-1L-mFlx}), the std values of revenues are much smaller under \textit{BiD} than  \textit{MyD}. However, with lower VRES amounts and higher generator flexibility, the std values under \textit{BiD} and \textit{MyD} are similar.

\textit{Revenue uplift for VRES: \textit{BiD} can effectively prevent  revenue loss for  VRES producers compared with \textit{MyD}.} 
Figure  \ref{fig:priceprofitcomp}c shows that under conditions of high VRES amounts and limited generator flexibility (e.g., \textit{40R-1L-mFlx} and \textit{70R-1L-mFlx}), wind producers experience negative profits under \textit{MyD}. This is due to high RTM prices and low DAM prices (in Figures \ref{fig:priceprofitcomp}a-b),  despite more scheduled VRES in the DAM.

\section{Conclusion}\label{section:conclusion}

In this paper, we propose a computationally efficient mechanism for adjusting the quantity of VRES in the DAM, which takes into account uncertain real-time imbalance costs while remaining compatible with the existing sequential market-clearing structure. To efficiently solve the bilevel optimization problems involved, we propose a linear relaxation technique based on strong duality and McCormick envelopes. This method can be applied to the relaxed unit-commitment problem for large-scale systems, where the conventional approach for bi-level optimization based on MILP fails. We evaluate our bilevel framework on the 1576-bus NYISO system and compare it with the myopic strategy where VRES producers bid the expected forecast value in the DAM. Numerical studies demonstrate that, in the future market with high penetration levels of VRES, our bilevel framework can achieve significant economic benefits by efficiently adjusting VRES dispatch in the DAM. For example, it can significantly reduce the expected system cost, and volatility of market prices and   revenues.

The proposed bilevel framework can provide a market-efficiency benchmark for VRES bidding quantities in the DAM. It can also potentially assist in market monitoring of VRES bidding strategies. In future work, we plan to explore decomposition methods to further improve computational efficiency and investigate how to effectively incorporate this framework into current electricity market designs.


\bibliography{storage}

\appendix

\begin{figure*}[t]
	\centering
	\hspace{-3ex}
	\subfigure[]{
		\raisebox{-1mm}{\includegraphics[width=2.4in]{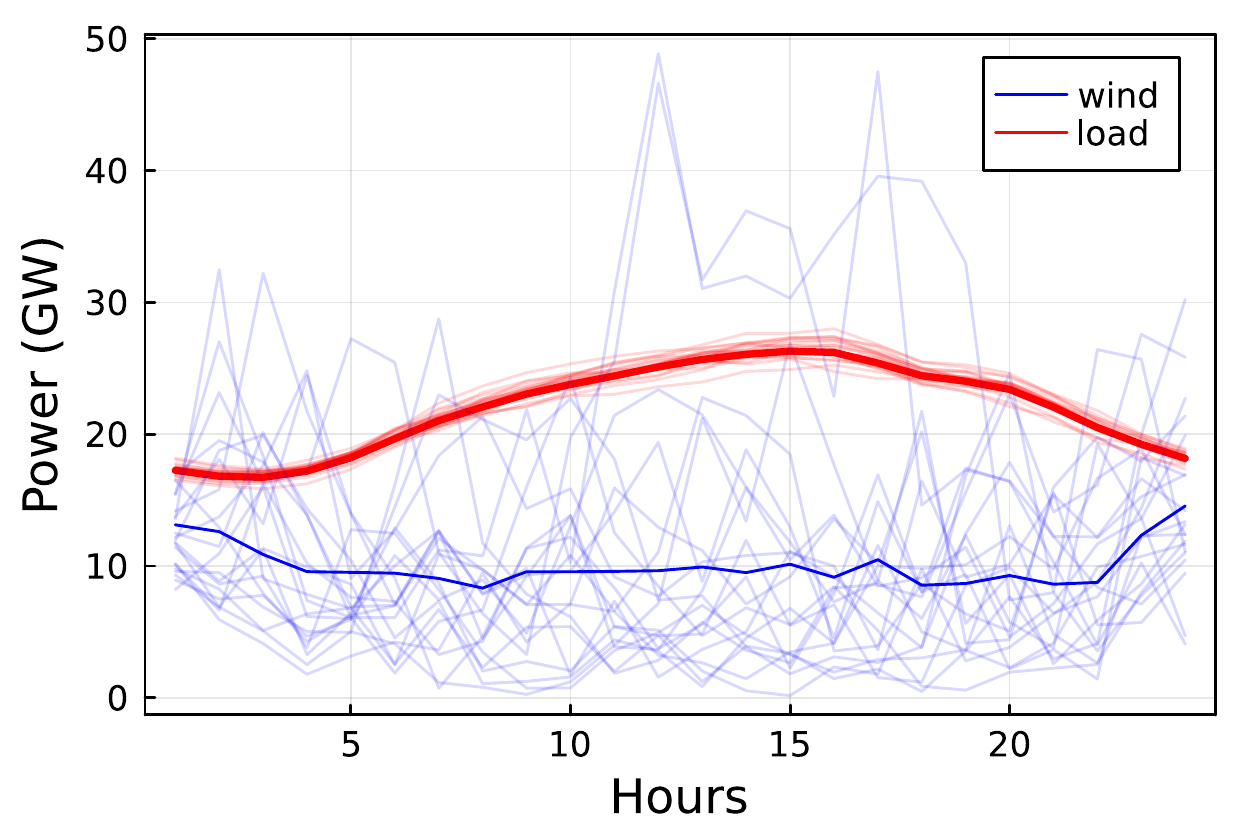}}}
	\hspace{-2ex}
	\subfigure[]{
		\raisebox{-1mm}{\includegraphics[width=2.4in]{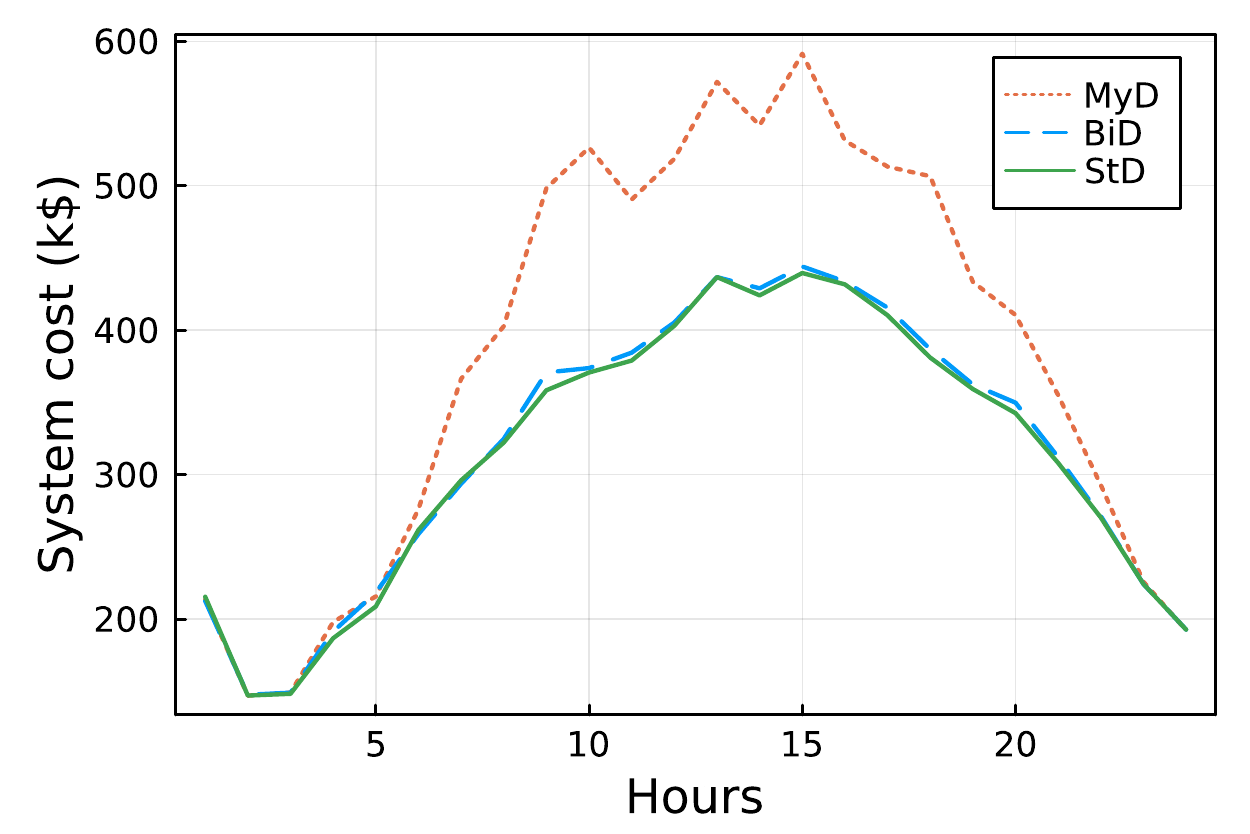}}}
         \hspace{-2ex}
	\subfigure[]{
		\raisebox{-1mm}{\includegraphics[width=2.4in]{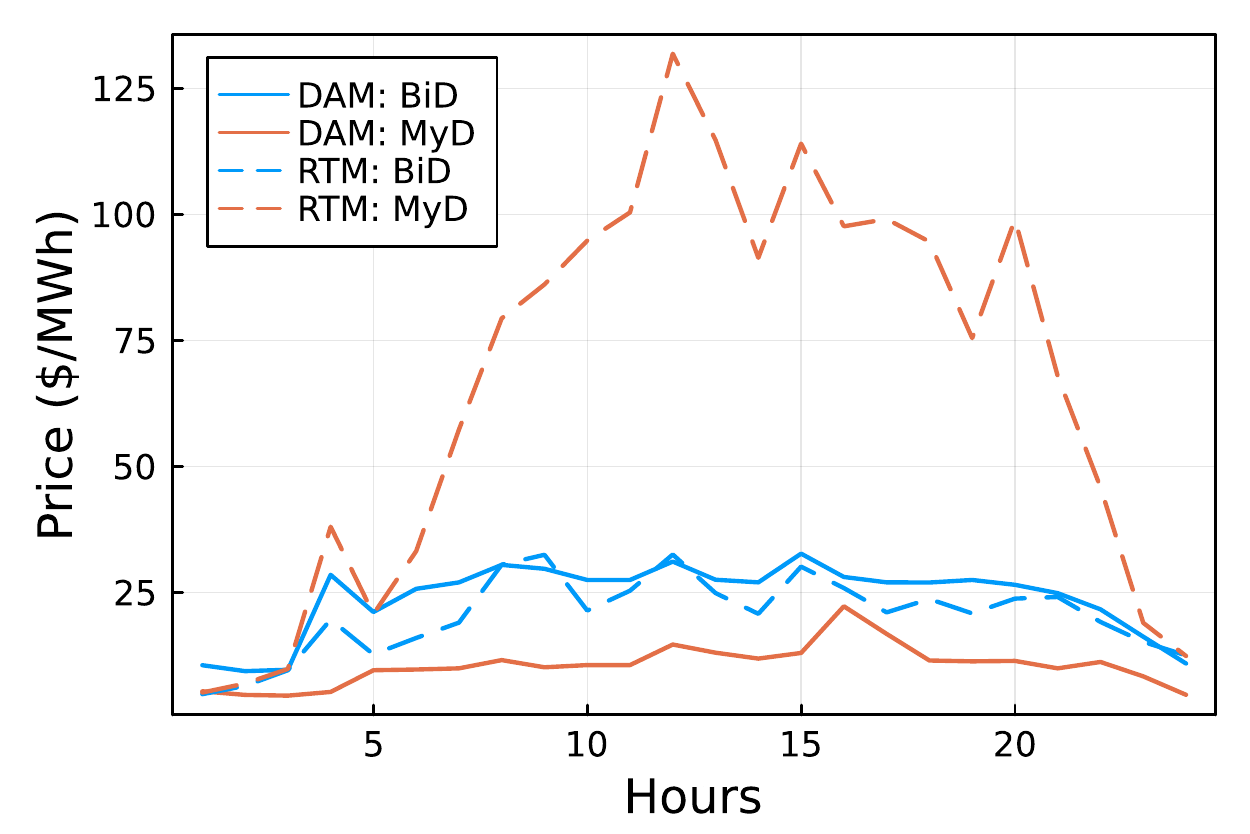}}}
	\vspace{-2mm}
	\caption{24-hour results: (a) Wind-energy generations and load; (b) Expected system costs; (c) Market prices. }
	\label{fig:day}
 \vspace{-2ex}
\end{figure*}

\subsection{24-hour Scheduling Horizon}
We present the result of  the 24-hour schedule of market-clearing models. Due to computation complexity, we solve the models of \textit{MyD},  \textit{BiD}, and \textit{StD}, respectively, for every 4-hour horizon,  and roll into 24 hours eventually. We use the parameter setting  \textit{40R-1L-mFlx} as one example.

In Figure \ref{fig:day}a, we show the scenarios of system wind energy (in blue) and load (in red) over 24 hours (on day August 2, 2019), where the highlighted curves show the average  profiles. In Figure \ref{fig:day}b, we present the 24-hour expected system costs for \textit{MyD} (in red),  \textit{BiD} (in blue), and \textit{StD} (in green), respectively. Figure \ref{fig:day}c simulates the DAM (solid curves) and RTM prices (dashed curves) for \textit{MyD} (in red) and  \textit{BiD} (in blue).

 As shown in Figure \ref{fig:day}, when the system load is high, e.g., around 15:00, the system cost under \textit{MyD} is much higher than that under \textit{BiD}. When the load is low, e.g., around 24:00,  the system costs under \textit{MyD} and \textit{BiD} are very close because we can have adequate flexible generators to deal with real-time uncertainty. In contrast, the system cost under \textit{BiD} is always close to \textit{StD}. Furthermore, in terms of market prices, RTM prices under \textit{MyD} can be very high in the mid-day while under \textit{BiD}, RTM and DAM prices are close.


\subsection{Performance of Relaxed Unit Commitment Problems}\label{app:uc}

In Section \ref{sec:market} of the main text, we relaxed binary variables of unit commitment into continuous variables $[0,1]$ to maintain the convexity of the lower-level problem. In this appendix, we show that this relaxation can well approximate the binary variable modeling. We focus on the setting of  \textit{40R-1L-mFlx}
.

 {First, we show that under the relaxation model, most of the unit-commitment solutions are either  0 or 1.} In Figure \ref{fig:uc}a, we show the frequency of relaxed DAM unit commitment solutions $u_{i,t}^{\text{DA}\star}$ (all hours and conventional generators) under \textit{BiD}. In Figure \ref{fig:uc}b, we show the frequency of relaxed RTM unit commitment solutions $u_{i,t,\omega}^{\text{RT}\star}$ (all hours, scenarios, and conventional generators) under \textit{BiD}. As shown in Figures \ref{fig:uc}a and  \ref{fig:uc}b, the solutions of 1 and 0 absolutely dominate all the solution outcomes. For DAM, the UC solutions lying in $(0,1)$ only make up 0.77\% of  the total number of  variables. For RTM, the UC solutions lying in $(0,1)$ only make up 0.89\%.

 {Second, we show that given the obtained bidding quantities of VRES, the relaxed model and the accurate binary model induce similar system costs. Although the bilevel model is based on relaxed unit-commitment decisions, the main purpose is to obtain the optimal bidding quantities of VRES. Hence, the system operator can continue to clear the DAM and RTM market using an accurate unit-commitment model with binary variables. To investigate this issue in more detail, we test the accurate unit-commitment model based on the optimal bidding quantities of VRES, which shows a system cost very similar to the relaxed unit-commitment model. Specifically, based on the relaxed unit-commitment model, we have the system costs (per hour) for NYISO: The DAM cost= 303 $K\$$;  the expected RTM cost= -3$K\$$; the total system cost= 300$K\$$. For the system costs based on the accurate unit-commitment model with binary variables, we have the DAM cost= 305$K\$$;  the expected RTM cost= -5$K\$$; the total system cost= 300$K\$$.}

\begin{figure}[t]
	\centering
	\hspace{-2ex}
	\subfigure[]{
		\raisebox{-1mm}{\includegraphics[width=1.77in]{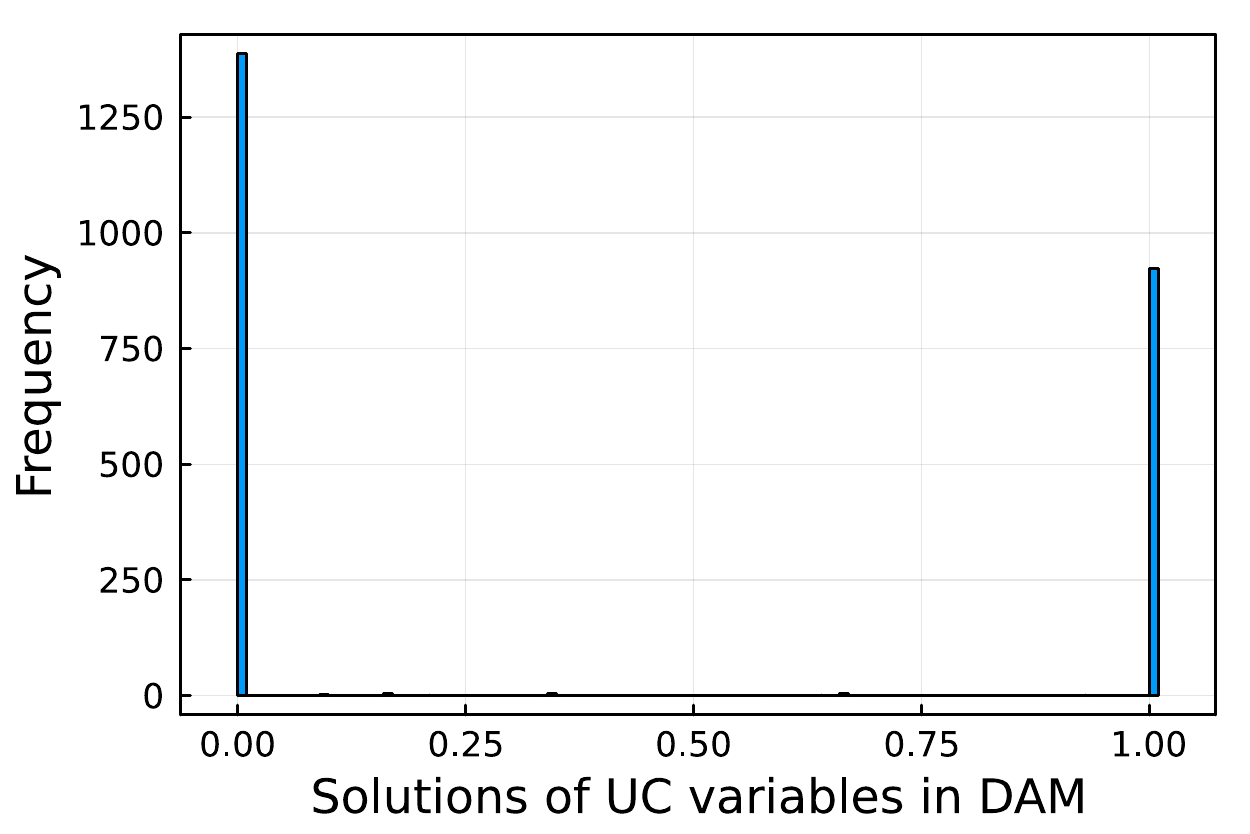}}}
	\hspace{-3ex}
	\subfigure[]{
		\raisebox{-1mm}{\includegraphics[width=1.77in]{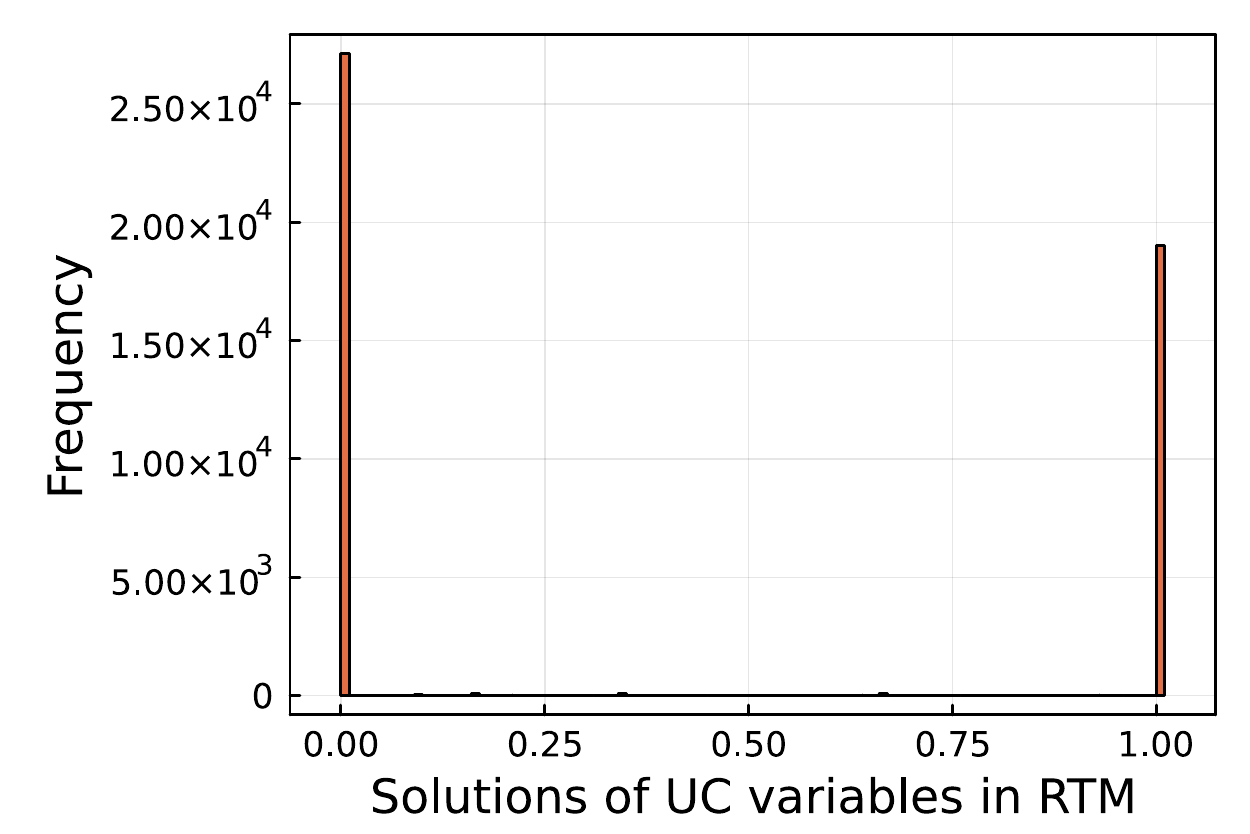}}}
	\vspace{-2mm}
	\caption{\textit{BiD}: (a) Frequency of UC solutions in DAM; (b) Frequency of UC solutions in RTM. }
	\label{fig:uc}
 \vspace{-2ex}
\end{figure} 

\endgroup

\subsection{Volatility Reduction} \label{app:volitility}

We show the volatility reduction for conventional generators' revenues and RTM re-dispatch amounts.

\begin{figure}[t]
	\centering
	\hspace{-3ex}
	\subfigure[]{
		\raisebox{-1mm}{\includegraphics[width=2.2in]{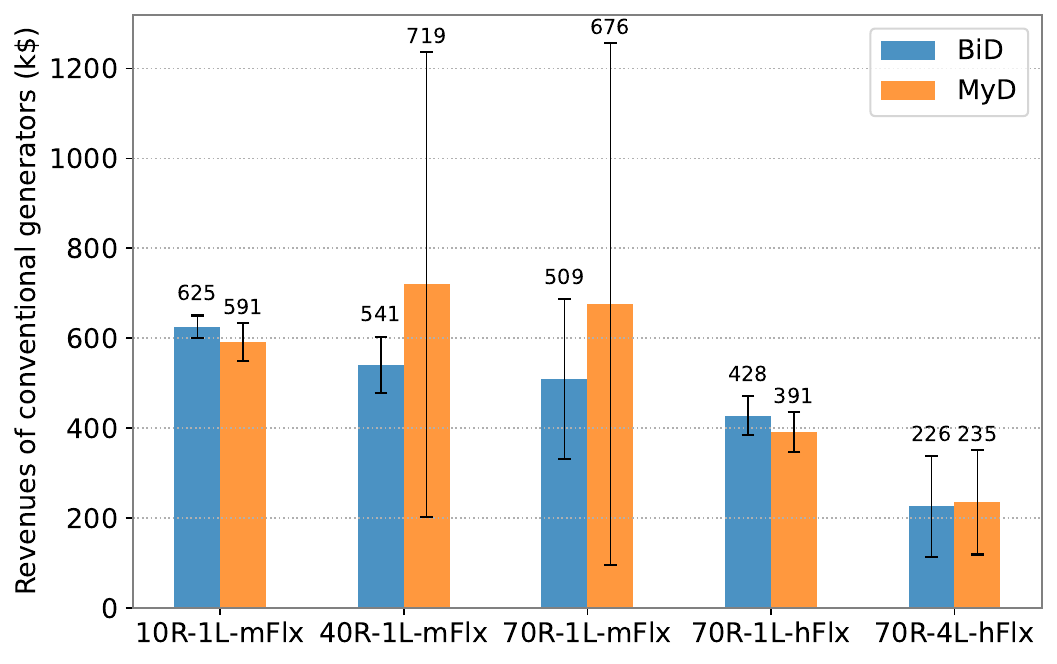}}}
         \hspace{-2ex}
	\subfigure[]{
		\raisebox{-1mm}{\includegraphics[width=2.2in]{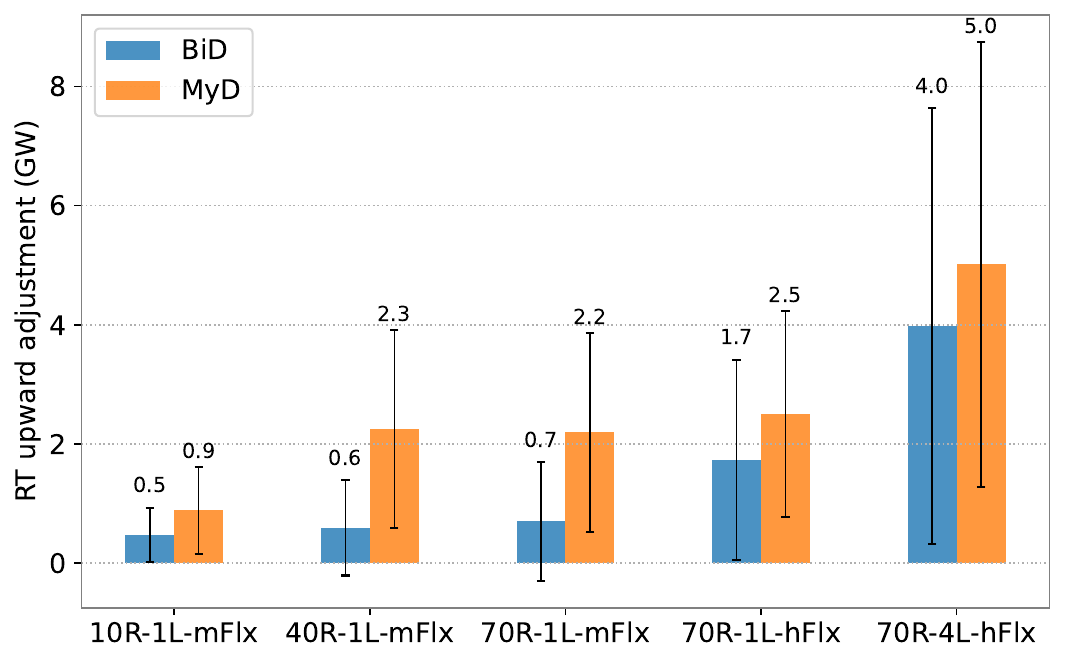}}}
	\vspace{-2mm}
	\caption{ \textit{BiD} vs \textit{MyD}: (a)  All conventional generators'  profits; (b) RTM upward adjustments. Both are average in hours. }
	\label{fig:barcomp}
 \vspace{-2ex}
\end{figure}

 In Figure  \ref{fig:barcomp}, under different system parameter settings, we show (a)  revenues of all conventional generators, and  (b) upward adjustments in RTM. The blue and orange bars show the results under \textit{BiD} and \textit{MyD}, respectively. We demonstrate the average values in terms of scenarios and hours, where the error bars show the std in scenarios.

\textit{Revenue volatility reduction: \textit{BiD} can reduce the volatility of  conventional generators' revenues  and  RTM upward adjustments.}   In Figures \ref{fig:barcomp}a-b, under high VRES amounts and limited generator flexibility (i.e., \textit{40R-1L-mFlx} and \textit{70R-1L-mFlx}), the std values of conventional units' revenues and RTM upward adjustments are much smaller under \textit{BiD} than under \textit{MyD}. However, with lower VRES amounts and higher generator flexibility, the std values under \textit{BiD} and \textit{MyD} are similar.

\textit{Upward-adjustment reduction:  Compared with \textit{MyD}, \textit{BiD} can significantly reduce the upward adjustment in the RTM.} As shown in Figure \ref{fig:barcomp}b,  \textit{MyD} requires higher upward adjustment in RTM, compared with \textit{BiD}, particularly under \textit{40R-1L-mFlx} and \textit{70R-1L-mFlx}. The reason is that with limited flexibility, \textit{BiD} tends to be more conservative in dispatching wind energy in DAM. Thus,
\textit{BiD} brings more conventional generators online, which mitigates the impact of VRES uncertainty in real time and reduces the requirement for upward adjustment.

\subsection{ {Upper Bound choices of Dual Variables in the McCormick Envelope}} \label{app:dual}

 {In Figure \ref{fig:bounds}, we show how the upper bounds of bidding quantities in the McCormick envelope impact the result. We now further demonstrate simulation results by varying the upper bound of the dual variables in the McCormick envelope. The results show that a lower upper bound can lead to infeasible solutions. However, when there are feasible solutions, the solutions are robust when we increase the upper bound of the dual variables. For demonstration, we use 5 scenarios to save computation time.}

\begin{figure}[t]
	\centering
	\hspace{-2ex}
	\subfigure[]{
		\raisebox{-2mm}{\includegraphics[width=1.72in]{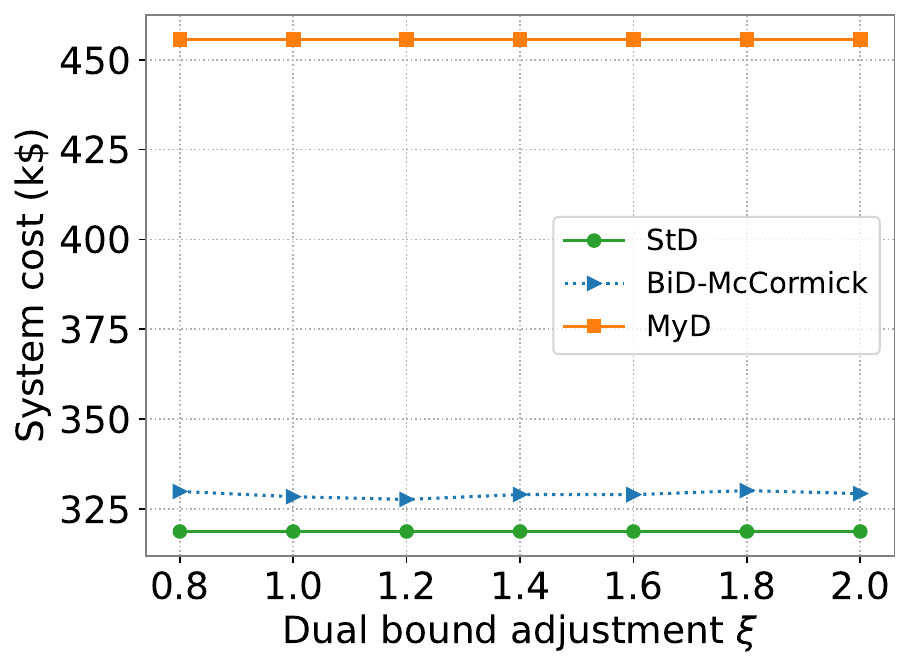}}}
	\hspace{-2ex}
	\subfigure[]{
		\raisebox{-2mm}{\includegraphics[width=1.72in]{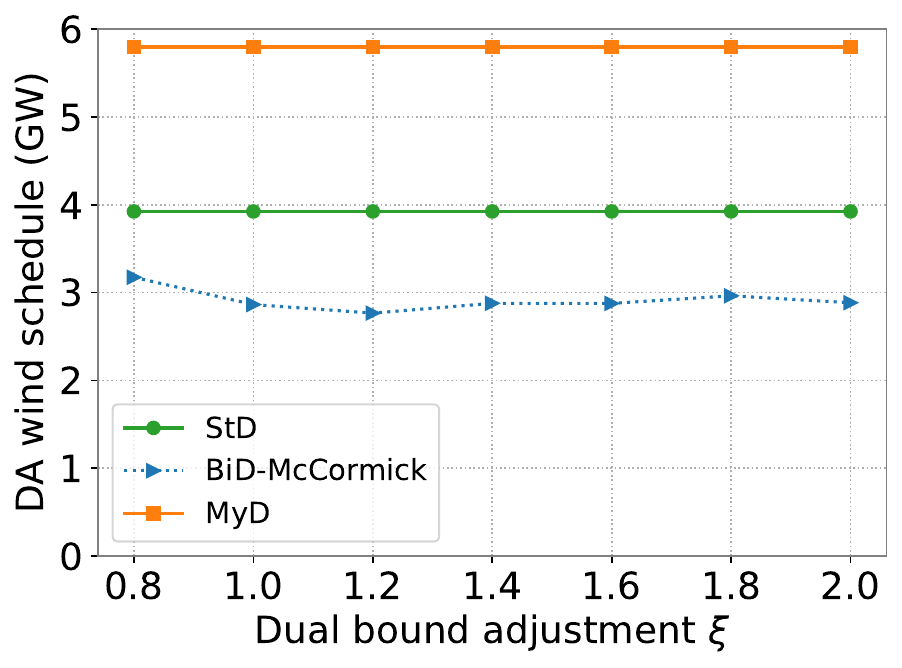}}}
	\vspace{-3mm}
	\caption{ (a) \small Expected system cost; (b) Day-ahead scheduled total VRES quantity. Both are functions of dual bound adjustment $\xi$.}
	\label{fig:bounds_dual1}
\end{figure}

 {Recall that we have the bilinear item $z_{k,t}=\overline{\lambda}_{k,t}^\text{W} \cdot W_{k,t}$. In  Figure \ref{fig:bounds}, we have discussed the impact of the upper bound of $W_{k,t}$. For the lower bounds and upper bounds of $\overline{\lambda}_{k,t}$, i.e., $0 \leqslant \overline{\lambda}_{k,t}^\text{W} \leqslant \beta_{k,t}^\lambda$, we set  the upper bound
$\beta_{k,t}^\lambda=\xi\cdot \overline{\lambda}^\text{W}_{k,t}(\bm{0})$, where $\overline{\lambda}^\text{W}_{k,t}(\bm{0})$ denotes the dual solution of $\overline{\lambda}^\text{W}_{k,t}$ in the DAM schedule when all VRES producers bid zero quantity. In the simulation results of Section \ref{section:simulation}, we let $\xi=1$. In this simulation, we vary $\xi$ between $[0,2]$. We find that when $0 \leq \xi \leq 0.6$, there is no feasible solution. However, when $0.8 \leq \xi \leq 2$, the solutions, e.g., system cost (Figure \ref{fig:bounds_dual1}a) and DAM wind schedule amount (Figure \ref{fig:bounds_dual1}b) will remain robust. This shows that the upper bound of the dual variables may have a limited impact on the solution accuracy while the upper bound of wind bidding quantities has a larger impact as shown in Figure \ref{fig:bounds}}.

\subsection{ {Impact of Number of Scenarios on Out-of-sample Test}} \label{app:scenario}

 {We demonstrate how the number of scenarios impacts the out-of-sample performance of the model.}

 {We test the cases of  10, 20, and 30 scenarios, respectively. We first obtain the optimal bidding quantities of VRES based on the 10, 20, and 30 scenarios, respectively. Then, in the out-of-sample test, for each case, we randomly generate 30 scenarios for VRES actual generations and examine the expected system cost under the newly generated 30 scenarios. We randomly generate 30 scenarios by 50 times to evaluate varieties.}
 
 { We show the mean value and the standard deviation (std) values among 50 tests for each case in Table \ref{tab:scenariox}.  A larger number of scenarios perform better in the out-of-sample test with lower expected system costs and lower std values. }

\begin{table}[bt!]
\caption{NYISO: mean and std over 50 tests.}
\vspace{-2ex}
\label{tab:scenariox}
\begin{center}
\begin{tabular}{lcccccccc}
\toprule
\# Scenarios  & mean (k\$) & std  (k\$)\\
\midrule
\textit{10}      & 310.00 &  6.73   \\
\textit{20} & 303.89  & 6.70   \\
\textit{30}   & 302.71 &  5.95    \\
\bottomrule
\end{tabular}
\end{center}
\vspace{-3ex}
\end{table}

\end{document}